\documentclass[galaxies,article,accept,pdftex,moreauthors]{Definitions/mdpi} 
\firstpage{1} 
\pubvolume{1}
\issuenum{1}
\articlenumber{0}
\pubyear{2025}
\copyrightyear{2025}
\externaleditor{Carlo Nipoti } 
\datereceived{28 February 2025} 
\daterevised{31 March 2025} 
\dateaccepted{05 April 2025 } 
\datepublished{ } 
\hreflink{https://doi.org/} 

\Title{Accurate Decomposition of Galaxies with Spiral Arms: Dust Properties and Distribution}

\TitleCitation{Accurate Decomposition of Galaxies with Spiral Arms: Dust Properties and Distribution}

\Author{Alexander A. Marchuk 
$^{1,2,}$*\orcidA{}, Ilia V. Chugunov $^{1}$\orcidC{}, Fr{\'e}d{\'e}ric Galliano$^{3}$\orcidG{}, Aleksandr V. Mosenkov $^{4}$\orcidB{}, \mbox{Polina V. Strekalova $^{1}$}, Valeria S. Kostiuk $^{5}$\orcidF{}, George A. Gontcharov $^{1}$\orcidD{}, Vladimir B. Il’in $^{1,2}$\orcidE{}, \mbox{Sergey S. Savchenko $^{1,2}$} 
, Anton A. Smirnov $^{1}$ and Denis M. Poliakov $^{1}$}

\AuthorNames{Alexander A. Marchuk, Ilia V. Chugunov, Fr{\'e}d{\'e}ric Galliano, Aleksandr V. Mosenkov, Polina V. Strekalova, Valeria S. Kostiuk, George A. Gontcharov, Vladimir B. Il’in, Sergey S. Savchenko, Anton A. Smirnov and Denis M. Poliakov}

\isAPAStyle{%
       \AuthorCitation{Lastname, F., Lastname, F., \& Lastname, F.}
         }{%
        \isChicagoStyle{%
        \AuthorCitation{Lastname, Firstname, Firstname Lastname, and Firstname Lastname.}
        }{
        \AuthorCitation{Marchuk
        , A.A.; Chugunov, I.V.; Galliano, F.; Mosenkov, A.V.; Strekalova, P.V.; Kostiuk, V.S.; Gontcharov, G.A.; Il’in, V.B.; Savchenko, S.S.; Smirnov, A.A.; et al.}
        }
}

\address{%
$^{1}$ \quad Central (Pulkovo) Astronomical Observatory, Russian Academy of Sciences, Pulkovskoye Chaussee 65/1, 196140 St.~Petersburg, Russia;\\ 
$^{2}$ \quad St. Petersburg State University
, 7/9 Universitetskaya Nab., 199034 St.Petersburg, Russia\\
$^{3}$ \quad Universit{\'e} Paris-Saclay, Universit{\'e} Paris Cit{\'e}, CEA, CNRS, AIM 
, 91191 Gif-sur-Yvette, France; \\
$^{4}$ \quad Department of Physics and Astronomy, N283 ESC, Brigham Young University, Provo, UT 84602, USA; 
$^{5}$ \quad Sternberg Astronomical Institute, Lomonosov Moscow State University, Universitetsky Pr. 13, \mbox{119234 Moscow, Russia};} 

\corres{Correspondence: a.a.marchuk+astro@gmail.com}

\abstract{We analyze three nearby spiral  galaxies---NGC~1097, NGC~1566, and NGC~3627---using images from the DustPedia database in seven infrared bands (3.6, 8, 24, 70, 100, 160, and 250~$\upmu$m). For each image, we perform photometric decomposition and construct a multi-component model, including a detailed representation of the spiral arms. Our results show that the light distribution is well described by an exponential disk and a S\'ersic bulge when non-axisymmetric components are properly taken into account. We test the predictions of the stationary density wave theory using the derived models in bands, tracing both old stars and recent star formation. Our findings suggest that the spiral arms in all three galaxies are unlikely to originate from stationary density waves. Additionally, we perform spectral energy distribution (SED) modeling using the hierarchical Bayesian code HerBIE, fitting individual components to derive dust properties. We find that spiral arms contain a significant {(>10\%)} fraction of cold dust, with an average temperature of approximately 18--20 K. The estimated fraction of polycyclic aromatic hydrocarbons (PAHs) declines significantly toward the galactic center but remains similar between the arm and interarm regions.}

\keyword{galaxies; spiral structure; galactic dynamics and kinematics; spiral arms; photometric decomposition; dust in galaxies} 

\begin{document}

\section{Introduction}

Dust plays a crucial role in the evolution of a galaxy, although~comprising less than 1\% of a galaxy's total mass. It absorbs and scatters starlight, then re-emits the absorbed energy in the far-infrared (FIR), causing thermal dust emission to dominate the spectral energy distribution (SED) at wavelengths longer than 10~$\upmu$m~\cite{2002MNRAS.335L..41P}. Dust is also a key tracer of the interstellar gas content, facilitating the formation of molecular hydrogen and other molecules by shielding them from ionizing radiation and acting as a catalyst for chemical reactions~\cite{1995ApJ...443..152W,1978ApJS...36..595D}. For~these reasons, dust is intricately linked to ongoing star formation, serving as both a tracer of star-forming regions~\cite{2007ApJ...666..870C,2009ApJ...703.1672K} and an active participant in the process by regulating the thermal balance of the interstellar medium (ISM) and contributing to molecule formation~\cite{2011ApJ...729L...3D}. Dust appears in various forms, including a diffuse component spread throughout the ISM~\cite{2000ASPC..197..237B}, dust lanes along spiral arms and bars~\cite{1999Ap&SS.269..423G}, and~dense, cold molecular clouds associated with star formation~\cite{2015A&A...579A..15K}. Thus, dust exhibits a vast range of physical properties that must be studied to fully understand its critical role in the formation and evolution of the~ISM.

With the ability to now measure dust properties in galaxies up to $z\sim7$ \cite{2025arXiv250110541F,2025arXiv250110508A}---and even farther thanks to ALMA and JWST---our understanding of dust characteristics in the local Universe has significantly advanced (see reviews in~\cite{2003ARA&A..41..241D,2018ARA&A..56..673G}). However, key knowledge gaps remain, including the fundamental question of dust grain composition and several unresolved issues in dust evolution and distribution. One major challenge is the dust energy balance problem, where observed FIR--submillimeter (submm) emission appreciably exceeds predictions based on observed dust attenuation~\cite{2000A&A...362..138P,2004A&A...425..109A,2010A&A...518L..39B,2012MNRAS.427.2797D,2016A&A...592A..71M,2018A&A...616A.120M}. This discrepancy suggests that current models of dust emissivity and radiative transfer (RT) may require revision. Another key issue relates to dust heating mechanisms. While the old stellar population has been found to play a significant role in heating dust~\cite{2012MNRAS.419.1833B,2019A&A...624A..80N,2020A&A...637A..25N}, the~efficiency of this process depends on the morphological component in which the dust resides, with~variations observed between different galaxy structures. Additionally, recent studies have revealed that large-scale dust distributions in galaxies are not purely exponential as~once assumed~\cite{2000A&A...353..117M,2009ApJ...701.1965M}. Instead, many galaxies exhibit significant central dust depletion, indicating a more complex dust structure~\cite{Mosenkov2019,2012MNRAS.423...38M}. These findings highlight the need for more detailed modeling of dust distribution and heating processes to fully understand the role of dust in galaxy~evolution.

These challenges have been addressed using a variety of advanced methods. One of the most effective techniques for examining dust distribution based on observed dust attenuation in edge-on galaxies is RT modeling. By~assuming a predefined geometry for each component~\cite{2017A&A...599A..64V,2019MNRAS.487.2753W,radiativeM81,radiativeM51,2020A&A...637A..25N} or by adopting an azimuthally averaged approach~\mbox{\cite{2020MNRAS.495..835T,2023MNRAS.526..118I}}, radiative transfer calculations can generate sophisticated models. These models are derived either from monochromatic images~\cite{1997A&A...325..135X,2007A&A...471..765B,2013A&A...550A..74D} or from multiple bands simultaneously, as~implemented in the Monte Carlo-based \verb|SKIRT| code~\cite{2011ApJS..196...22B,2015A&C.....9...20C}.  

The evolution of dust content in galaxies has been explored via SED fitting at both unresolved~\cite{2021A&A...649A..18G,2023AJ....165..260D,2010A&A...518L..61B} and, less frequently, resolved~\cite{2012ApJ...756..138A,2025ApJS..276....2C,2022ApJ...926...81A,2024A&A...687A.255S,2024ApJ...973..137S} scales. Various tools, including \verb|CIGALE| \cite{2019A&A...622A.103B}, \verb|piXedfit| \cite{2022ApJ...926...81A}, \verb|MAGPHYS| \cite{2008MNRAS.388.1595D}, \verb|FAST| \cite{2009ApJ...700..221K}, and \verb|HerBIE| \cite{2018MNRAS.476.1445G}, along with modern dust grain models such as \verb|THEMIS| \cite{2017A&A...602A..46J} and the `astrodust'-based model~\cite{2023ApJ...948...55H}, have facilitated the study of the relationship between dust content and key galactic properties, including gas abundance, metallicity, stellar mass, and~star formation rate~\cite{2010MNRAS.403.1894D,2012A&A...540A..52C,2021A&A...649A..18G,2019A&A...624A..80N}.  

The photometric decomposition of individual galactic subsystems, such as the central bulge or the extended disk, enables a detailed examination of dust and stellar distributions on galactic scales, providing insights into their geometry and spatial distribution. This analysis is typically performed using tools like \verb|IMFIT| \cite{Erwin2015} or \verb|GALFIT| \cite{2010AJ....139.2097P}, which employ standard parametric functions, such as the S{\'e}rsic function for the bulge and an exponential function for the disk. {At the same time, the~presence of dust hampers photometric analysis because it is not taken into account in a self-consistent manner, and~needs manual masking or techniques that work in a limited number of situations~\cite{2023MNRAS.524.4729S}.} Nevertheless, photometric decomposition has proven to be a valuable technique in both RT~\cite{radiativeM81} and SED~\cite{2023ApJS..267...26G} modeling, offering crucial insights into the interplay between dust and stellar components in~galaxies.

Spiral galaxies are the most numerous type of non-dwarf galaxy in the local Universe~\cite{2006MNRAS.373.1389C,2013MNRAS.435.2835W} and remain abundant even at high redshifts (\cite{Chugunov2025,2024ApJ...968L..15K}, and also~\cite{Salcedo2025}). Spiral arms, which emit a significant fraction of galactic light~\cite{Savchenko2020,Marchuk_M51,Chugunov2024}, are optically thick in contrast to the optically thin interarm regions~\cite{2000ApJ...542..761W} and play a crucial role in shaping the chemical composition of galaxies on large scales~\cite{2018A&A...611L...2K,2013MNRAS.428..625S}.
However, their proper inclusion in RT and SED models remains~rare. 

In RT models, spiral arms are often approximated by an additional disk component~\mbox{\cite{2009EAS....34..247P,2023MNRAS.526..118I}}, with~some exceptions~\cite{2000A&A...353..117M,2012ApJ...746...70S}, where authors have demonstrated a clear preference for models that explicitly incorporate a spiral structure. Given that the star-to-dust geometry has been shown to be a critical factor in RT modeling~\cite{2025arXiv250112008G}, properly accounting for spiral arms is essential. It is therefore unsurprising that existing models that omit this attribution often deviate from observed images, particularly in spiral arms and features associated with them, such as ``bumps'' in azimuthal profiles~\cite{2020A&A...637A..25N,2022MNRAS.514..113R,2025MNRAS.537...56P}.  

In SED fitting, dust properties in spiral arms can be studied using resolved maps, but~such studies are generally limited to a small number of objects due to the significant computational effort required. Additionally, arms must still be properly delineated~\cite{2024ApJ...973..137S,2024A&A...687A.293Q}. Alternatively, models can be constructed for individual galactic components~\cite{2023ApJS..267...26G,2005A&A...441..491V} but, to~our knowledge, such an approach has never been applied specifically to spiral arms. This omission is largely due to the difficulty of accurately modeling spiral arms in photometric decomposition studies, {where, so far, only the \verb|GALFIT| approach with Fourier and bending modes, modified by a rotation function, produces some adequately modeled arms} \cite{2010AJ....139.2097P,2017ApJ...845..114G}.  

In the series of papers~\cite{Chugunov2024,Marchuk_M51,Chugunov2025}, we addressed these issues and introduced a new photometric model that represents individual spiral arms. The~model is highly flexible and has been refined throughout our investigation, allowing for significant variation in both arm shape and the light distribution along and across the~arm.  

In the first paper of this series, the~model was applied to 29 galaxies using 3.6~$\upmu$m band images~\cite{2015ApJS..219....4S}, revealing a connection between the pitch angle and bulge/bar fraction, as~well as an increase in the spiral-to-total light ratio in more luminous disks. In~the second paper~\cite{Marchuk_M51}, we focused on the iconic galaxy M\,51, utilizing a wide wavelength range from far-ultraviolet (FUV) to far-infrared, with~a total of 17 images from the DustPedia project~\cite{dustpedia}. By~tracing different stellar populations, we tested for the presence of angular offsets associated with the density wave theory~\cite{Miller2019}, examined how spiral pattern properties vary with wavelength, and~highlighted the importance of an accurate bulge and disk model.  In~M~51, we do not observe the signs of a long-lived density wave in the whole spiral~pattern.

In the final paper of the series~\cite{Chugunov2025}, we applied the model to a significantly larger sample of 159 galaxies, using data from the HST COSMOS~\cite{2007ApJS..172..196K}, JWST CEERS~\cite{Bagley2023}, and~JADES~\cite{2024A&A...690A.288B} surveys. Spanning a wide redshift range of \( 0.1 \leq z \leq 3.3 \), the~model was applied to images in optical and near-infrared rest-frame wavelengths. Our results indicate significant evolution in spiral arm properties over cosmic time. Specifically, the~pitch angle decreases at a rate of 0.5 deg/Gyr (see also~\cite{2022AstL...48..644R,2023A&A...680L..14R}) up to 11 Gyr (corresponding to \( z \approx 2.5 \)), while the spiral-to-total ratio exhibits a slight increase with lookback~time.  

Overall, our findings in~\cite{Chugunov2024,Marchuk_M51,Chugunov2025} demonstrate that the proposed model can successfully be applied to construct a broad range of spiral arm structures with diverse morphologies and across different~wavelengths.

The numerous challenges in modeling spiral arms stem from the lack of a conclusive understanding of their nature, even within our own galaxy~\cite{2024MNRAS.533.4324F}. While advances in the quasi-stationary density wave theory~\cite{Lin1964,Lin1967,Roberts1969,Bertin1989} address the so-called ``winding dilemma'', demonstrating this theory in N-body simulations remains challenging.  
The influence of bars~\cite{2012MNRAS.426L..46A,2007A&A...472...63R} and tidal interactions caused by close flybys~\cite{1969ApJ...158..899T} can generate spiral patterns, but~these structures may not be sustained over long periods. Alternatively, spiral arms may have a ``dynamic'' nature~\cite{Julian1966,Sellwood1984,Sellwood2011}, co-rotating with the disk and forming through recurrent cycles of groove instabilities~\cite{2019MNRAS.489..116S,Dobbs2010} or episodic stochastic star formation.  
A comprehensive review of these topics and competing theories can be found in~\cite{2021arXiv211005615S,Dobbs2014,2016ARA&A..54..667S}.

Observations in IR bands provide valuable insights into the nature of spiral arms. Measurements in the near-infrared (NIR) bands can be used to estimate stellar mass and gravitational potential~\cite{2015ApJS..219....4S}, while observations in the mid-infrared (MIR) filters serve as reliable tracers of the star formation rate (SFR) \cite{2007ApJ...666..870C,Tamburro2008,2015ApJS..219....8C,2009ApJ...703.1672K}.  
FIR emission can also be used to measure the SFR and provides a means to estimate cold dust mass~\cite{1990ApJ...350L..25D,2012MNRAS.425..763G,2013MNRAS.428.1880A,2014MNRAS.440..942C}. Under~certain assumptions, dust mass estimates can be converted into gas mass fractions~\cite{2021A&A...649A..18G,2022ApJ...926...81A,2025ApJS..276....2C}, offering further insights into the ISM properties of spiral~galaxies.

To summarize, there are still plenty of questions about the dust---its distribution,  mass, composition, and temperature on the galactic scale---and, from~the dynamical perspective, we still face questions about the nature of large nonaxisymmetric features, such as spiral arms. At~the same time, the proper photometric decomposition of resolved galaxies in several IR bands simultaneously could be quite beneficial for helping with these issue for the reasons listed above. In~this paper, we decompose the images of three nearby spiral galaxies observed in the {\it Herschel} and {\it Spitzer} bands, covering a wavelength range from 3.6~$\upmu$m to 250~$\upmu$m. We model the classical bulge, bar, and~disk components, along with the prominent spiral arms, using the previously developed sophisticated model~\cite{Chugunov2024,Marchuk_M51,Chugunov2025}. The~broad wavelength coverage enables us to  
(i) investigate the dust distribution in these galaxies,  
(ii) test for the presence of a density wave, and~ 
(iii) perform SED modeling of individual galactic components, estimating key properties such as dust mass ($M_{\mathrm{dust}}$) and dust temperature~($T_{\mathrm{dust}}$).

Our paper is organized as follows. In Section~\ref{sec:data}, we describe the sample selection process and the galaxies and images used in this study. Section~\ref{sec:methods} outlines the details of the decomposition process (Section~\ref{sec:decomposition}) and the SED fitting with \verb|HerBIE| (Section~\ref{sec:sedfitting}).  In Section~\ref{sec:results}, we present the main results, including the light distribution analysis (Section~\ref{sec:decomp_results}), an~evaluation of the nature of spiral arms (Section~\ref{sec:cr}), and~the dust properties of individual galactic components (Section~\ref{sec:sedresults}). Finally, we summarize our findings in Section~\ref{sec:conclusion}.


\section{Data}
\label{sec:data}

As in our previous study of M\,51 \citep{Marchuk_M51}, we use the DustPedia
\endnote{\url{http://dustpedia.astro.noa.gr/} (accessed on 10 06 2024)} \citep{2017PASP..129d4102D} project as the primary data source. DustPedia, which is based on objects observed by the {\it Herschel} Space Observatory \citep{2010A&A...518L...1P}, is well suited for studying the role of dust in galaxies, making it an ideal fit for this~work.  

The database also includes a wide range of photometric data from other infrared surveys, such as the Two Micron All-Sky Survey (2MASS;~\cite{2006AJ....131.1163S}), the~Wide-field Infrared Survey Explorer (WISE;~\cite{2010AJ....140.1868W}), the~{\it Spitzer} Space Telescope \citep{2004ApJS..154....1W}, the~InfraRed Astronomical Satellite (IRAS;~\cite{1984ApJ...278L...1N}), and~the {\it Planck} Observatory \citep{2011A&A...536A...1P}. Additionally, DustPedia incorporates data from other parts of the spectrum, including the GALaxy Evolution eXplorer (GALEX;~\cite{2007ApJS..173..682M}), the~Sloan Digital Sky Survey (SDSS;~\cite{2000AJ....120.1579Y}), and~the Digitized Sky Survey (DSS\endnote{\url{https://archive.eso.org/dss/dss} (accessed on 10 06 2024)}), among~others.  

Images in all photometric bands have been carefully processed by~\cite{dustpedia}, ensuring convenient and user-friendly access to the required~data.

The multiwavelength catalog presented in DustPedia contains photometric data for 875 galaxies in the local Universe \citep{dustpedia}. We visually selected galaxies from the catalog, choosing those that were not too small, not excessively inclined relative to the line of sight (\(i < 70^{\circ}\)), and~exhibited clearly visible, prominent spiral arms.  The~preliminary sample, based on visual inspection, included 26 galaxies that display a variety of spiral patterns and angular sizes ranging from 2~arcmin to 20~arcmin. Three of these galaxies (NGC~4165, NGC~5364, and~NGC~5427) were previously analyzed in a single 3.6~$\upmu$m band in our earlier study~\cite{Chugunov2024}. Before~proceeding with further analysis, all images were rotated to a ``face-on'' orientation with respect to the line of sight, using orientation parameters from~\cite{Mosenkov2019} (Appendix~D). Additionally, the~sky background was subtracted where~necessary.

The goal of this study is to analyze the distribution and properties of different types of dust (cold, warm, and~hot) in galactic disks. To~achieve this, we require the broadest possible wavelength coverage from the NIR to the FIR. However, longer wavelengths result in lower spatial resolution and a larger point spread function (PSF) full width at half maximum (FWHM). Since we aim to investigate SEDs and reliably compare the properties of spiral structures in images obtained with different telescopes, all images must be processed to a uniform resolution and PSF. Consequently, the~choice of the longest wavelength used influences the entire dataset. We adopt the SPIRE 250~$\upmu$m band as the limiting wavelength for several reasons:  
(1) it provides an optimal balance between resolution and sensitivity;  
(2) for $\lambda > 250$~$\upmu$m, image resolution rapidly deteriorates; and  
(3) it is close to the peak of cold dust emission, making it well suited for studying dust properties. For~these reasons, we \citep{Marchuk_M51} and other researchers \citep{2024ApJ...973..137S} have used SPIRE 250~$\upmu$m as the limiting wavelength in similar~studies.  

In total, for~each galaxy, we collected images in seven filters: 3.6~$\upmu$m, 8~$\upmu$m, 24~$\upmu$m, 70~$\upmu$m, 100~$\upmu$m, 160~$\upmu$m, and~250~$\upmu$m. These images were obtained using the {\it Herschel} Space Observatory, with~the Spectral and Photometric Imaging REceiver (SPIRE;~\cite{2010A&A...518L...3G}) and the Photodetector Array Camera and Spectrometer (PACS;~\cite{2010A&A...518L...2P}), as~well as the {\it Spitzer} Space Telescope, using the InfraRed Array Camera (IRAC;~\cite{2004ApJS..154...10F}). The~filters and their parameters are listed in Table~\ref{tab:bands}.  {We chose to rely on Spitzer observations rather than those from WISE for constraining dust properties in our galaxy sample. The~WISE W1 (3.4~$\upmu$m) and W2 (4.6~$\upmu$m) bands are either nearly equivalent to Spitzer IRAC1 (3.6~$\upmu$m) and IRAC2 (4.5~$\upmu$m) bands or do not intersect significantly with key PAH emission features, offering limited additional diagnostic value. Meanwhile, the~W3 (12~$\upmu$m) and W4 (22~$\upmu$m) bands overlap with Spitzer MIPS coverage, but~the Spitzer data provide superior sensitivity, resolution, and~calibration. Additionally, the~WISE W3 and W4 channels were affected by cryogenic limitations, especially after coolant depletion, which compromised data quality in these bands. Given these considerations, Spitzer data offer more reliable constraints on mid- and far-infrared dust emission.}

\begin{table}[H] 
\caption{Infrared bands used in this work. Pixel size and PSF FWHM are given in~arcseconds.\label{tab:bands}}
\begin{tabularx}{\textwidth}{lCCCC}
\toprule
\textbf{Name/Facility} & \textbf{\boldmath$\lambda$, \boldmath$\upmu$m}  & \textbf{Pixel Size} & \textbf{PSF FWHM} & \boldmath$\log_{10}(\lambda [\upmu\mathrm{m}])$ \\
\midrule
Spitzer 3.6~$\upmu$m & 3.6 & 0.75 & 1.66 & 0.56 \\
Spitzer 8.0~$\upmu$m & 8.0 & 0.6 & 1.98 & 0.90 \\
Spitzer 24~$\upmu$m & 24 & 1.5 & 6 & 1.38 \\
PACS 70~$\upmu$m & 70 & 4 & 18 & 1.85 \\
PACS 100~$\upmu$m & 100 & 3 & 10 & 2.00 \\
PACS 160~$\upmu$m & 160 & 4 & 13 & 2.20 \\
SPIRE 250~$\upmu$m & 250 & 6 & 18 & 2.40 \\
\bottomrule
\end{tabularx}
\end{table}

Since we aim to construct SEDs and enable a robust comparison of parameters across individual morphological components, all images must be convolved to match the pixel scale and FWHM of the SPIRE 250~$\upmu$m band, which are 6~arcsec and 18~arcsec, respectively. This convolution was performed using transition kernels from \citet{2011PASP..123.1218A} and the PSF matching tools from {\small PHOTUTILS} \citep{2020zndo...4044744B}.

After the convolution step, approximately half of the galaxies in the preliminary sample appeared as too small and were excessively ``blurred'' by the PSF, making it difficult to distinguish individual features. Further selection among the remaining galaxies revealed several issues. In~some cases, such as IC~342 and NGC~3031, the~images were contaminated by bright galactic cirrus, which, despite its distinct color and geometric properties \citep{2020A&A...644A..42R,2021MNRAS.508.5825M,2023MNRAS.519.4735S}, is challenging to separate from the underlying galaxy. Additional complications arose due to framing issues (e.g., NGC~3184) or the presence of a nearby bright source (e.g., NGC~4725). Moreover, in~a significant number of galaxies, the~dust disk appeared as highly noisy, and~the spiral arms were flocculent, as~observed in NGC~5236, NGC~5457, and~NGC~6946. {Since flocculent galaxies have many hardly resolved arms}, constructing a reliable photometric model is either infeasible or requires substantial manual effort. As~a result, we selected three galaxies from the initial sample---NGC~1097, NGC~1566, and~NGC~3627---for which our photometric modeling appeared to be successful, as~described in the following~sections.










NGC~1097 is a strongly barred spiral galaxy, typically classified as SB(s)b, and~located at a distance of 13--15~Mpc~\cite{2021MNRAS.501.3621A,2001ApJ...553...47F}. Morphologically, it features a pair of clearly defined spiral arms that converge into two prominent dust lanes extending along the bar for approximately 10~kpc. Additionally, its spiral structure exhibits visible tidal distortions associated with strong interactions with its companion galaxy, NGC~1097A, located to the northwest. Within~the central region, at~a radius of $\sim$700~pc (8--10~arcsec), there is a ring of intense star formation~\cite{1995AJ....110..156Q,2023MNRAS.524..207K}, with~a current SFR of $\sim 1.8 \,\mathrm{M_{\odot}/yr}$ \cite{2019MNRAS.485.3264P}. It is suggested that an instantaneous burst of star formation occurred inside the ring approximately 7 million years ago~\cite{2000A&A...353..834K}, leading to the formation of super star clusters visible in the optical range~\cite{1995AJ....110.1009B}. NGC~1097 also hosts an active galactic nucleus (AGN), exhibiting optical luminosity variations and transitioning between LINER 1 and Seyfert 1 classifications~\cite{1995ApJ...443..617S,2008ARA&A..46..475H}. The~nucleus contains a supermassive black hole (SMBH) with a mass of approximately $\sim1.2\times10^8\, \mathrm{M_{\odot}}$ \cite{2006ApJ...642..711L}.



NGC~1566 is a barred spiral galaxy classified as SAB(s)bc and is the brightest member of the Dorado subgroup. It is so conspicuous in the southern celestial hemisphere that it has earned the nickname ``The Spanish Dancer''. {The distance estimates ranging from 6.5~Mpc to 17.8~Mpc according to different sources~\cite{2021MNRAS.501.3621A,2022MNRAS.514..403T}, and~we adopt the latter since it was obtained with the more precise TRGB method.} NGC~1566 is best known for its changing-look active galactic nucleus (CL AGN), which exhibits significant periodic flux variations across the optical, X-ray, and~UV spectra~\cite{2019MNRAS.483..558O,2022JHEAp..33...20L}, likely driven by nuclear disk instabilities~\cite{2019MNRAS.483L..88P}. The~galaxy hosts an SMBH with an estimated mass of $\sim10^7\, \mathrm{M_{\odot}}$ \cite{2002ApJ...579..530W}. Some studies, however, suggest the presence of two merging SMBHs~\cite{2021MNRAS.507..687J}.  
NGC~1566 features two remarkably prominent spiral arms that extend through multiple revolutions and are clearly visible in both the stellar and gaseous disks~\cite{2021A&A...656A.133Q,2024A&A...691A.163E,2024A&A...690A..69M}.

NGC~3627 (M\,66) is classified as an SAB(s)b-type galaxy~\cite{1991rc3..book.....D} with a prominent bar marked by luminous tips. It is a member of the Leo Triplet, which also includes NGC~3623 and NGC~3628. A~strong tidal interaction is observed between NGC~3628 and NGC~3627, manifesting as an asymmetry in the gas and dust distribution, while NGC~3623 appears as largely unaffected~\cite{1979ApJ...229...83H,2001A&A...378...40S}. NGC~3627 hosts a low-luminosity AGN of Seyfert 2 type~\cite{Moustakas2010}. Due to its distinct characteristics, it has been the focus of numerous studies, including but not limited to~\cite{2011A&A...527A..92C,2024ApJ...968...87K,2024MNRAS.531..815L}. Additionally, NGC~3627 has hosted multiple recent supernovae, with~at least three recorded in the past 30~years.  



To conclude this section, the~final sample consists of three late-type Sb galaxies: NGC~1097, NGC~1566, and~NGC~3627. Their properties are listed in Table~\ref{tab:galdata}. In~all cases, these galaxies exhibit at least two well-defined spiral arms in the IR bands, enabling their~decomposition.  

A strong bar is clearly visible in NGC~1097 and NGC~3627, while NGC~1566 also hosts a bar, though~significantly smaller. Another common feature among these galaxies is that they all harbor AGNs and belong to galaxy groups. Additionally, all three are included in the PHANGS-ALMA observational program~\cite{2021ApJS..255...19L}, which provides available spiral masks~\cite{2024A&A...687A.293Q}.  

Despite these similarities, the~galaxies represent fundamentally different cases in terms of dust distribution. According to the analysis by~\cite{Mosenkov2019}, their S{\'e}rsic profiles differ significantly:  
NGC~1097 has a very large S{\'e}rsic index (\(n \gg 1\)), 
NGC~1566 exhibits an intermediate value (\(n \sim 1\)), and~ 
NGC~3627 has a near-zero S{\'e}rsic index (\(n \sim 0\)).  

\begin{figure}[H]
\includegraphics[width=10.5 cm]{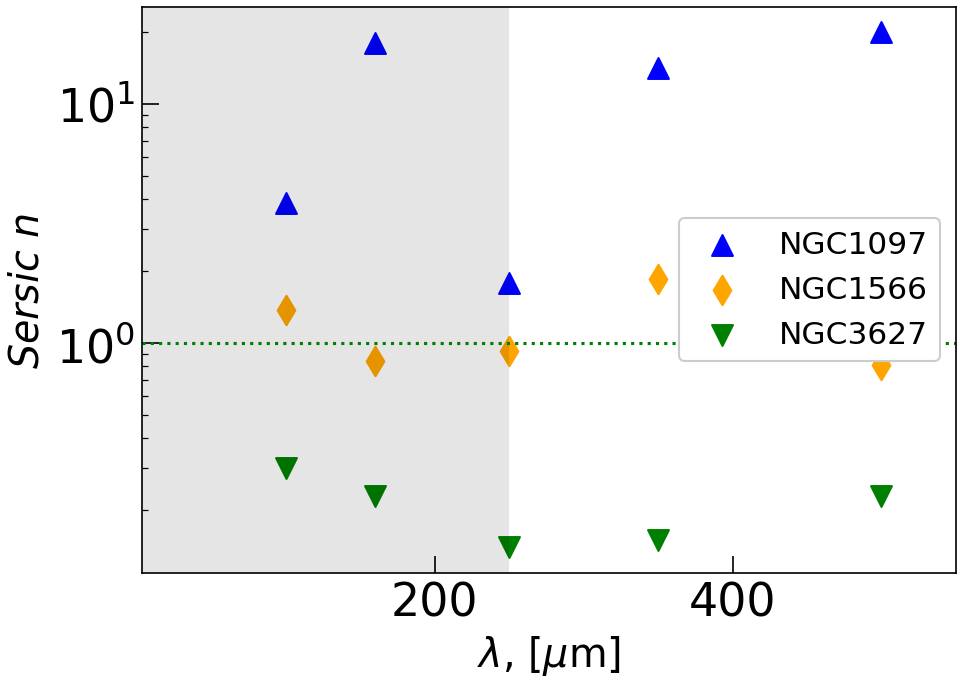}
\caption{The S{\'e}rsic indices, $n$, for~the galaxies under investigation, as~reported by~\cite{Mosenkov2019}. The~light-gray-filled area marks the wavelength range studied in this work.
 \label{fig:sers}}
\end{figure}

These differences, illustrated in Figure~\ref{fig:sers}, make these galaxies ideal candidates for testing more sophisticated decomposition~models.

\unskip

\begin{table}[H] 
\caption{Information about the galaxies in the sample, taken from~\cite{Makarov2014}. Distances were obtained from the compilation in Table~A2 of~\cite{2021MNRAS.501.3621A}.
 \label{tab:galdata}}
\begin{tabularx}{\textwidth}{lCCC}
\toprule
{} & \textbf{NGC~~1097} & \textbf{NGC~1566} & \textbf{NGC~3627} \\
\midrule
R.A. (J2000)   & $\text{2}^h\text{46}^m\text{19.12}^s$ &   $\text{4}^h\text{20}^m\text{00.42}^s$ & $\text{11}^h\text{20}^m\text{15.02}^s$      \\
D.A. (J2000)   & $-\text{30}^{\circ}$16$'$30$''$
          & $-\text{54}^{\circ}$56$'$16$''$ & $\text{+12}^{\circ}$59$'$30$''$        \\
Type  & SBb                                     &   SABb & Sb     \\
$i$            & $\text{54.8}^{\circ}$                      &   $\text{49.1}^{\circ}$ & $\text{67.5}^{\circ}$     \\
P.A.           & $\text{134}^{\circ}$                     &   $\text{44}^{\circ}$ & $\text{168}^{\circ}$      \\
Distance     & $13.6\pm2.0$~Mpc                                  &     $17.7\pm2.0$~Mpc  & $11.3\pm0.5$~Mpc    \\
Linear scale     & 394~pc/pix                                 &     513~pc/pix  & 501~pc/pix    \\
<$B-V$>        & 0.68~mag                             &   0.57~mag  & 0.63~mag      \\
$B$ magnitude        & 9.69~mag                             &   9.97~mag  & 9.09~mag      \\
$r_{25}$       & 314~arcsec     &   217~arcsec & 307~arcsec  \\  
\bottomrule
\end{tabularx}
\end{table}

\section{Methodology}
\label{sec:methods}

In this section, we provide a detailed description of the photometric decomposition process (Section~\ref{sec:decomposition}) and the subsequent SED modeling of individual components (\mbox{Section~\ref{sec:sedfitting}}).

\subsection{Decomposition}
\label{sec:decomposition}

\begin{figure}[H]
\includegraphics[width=10cm]{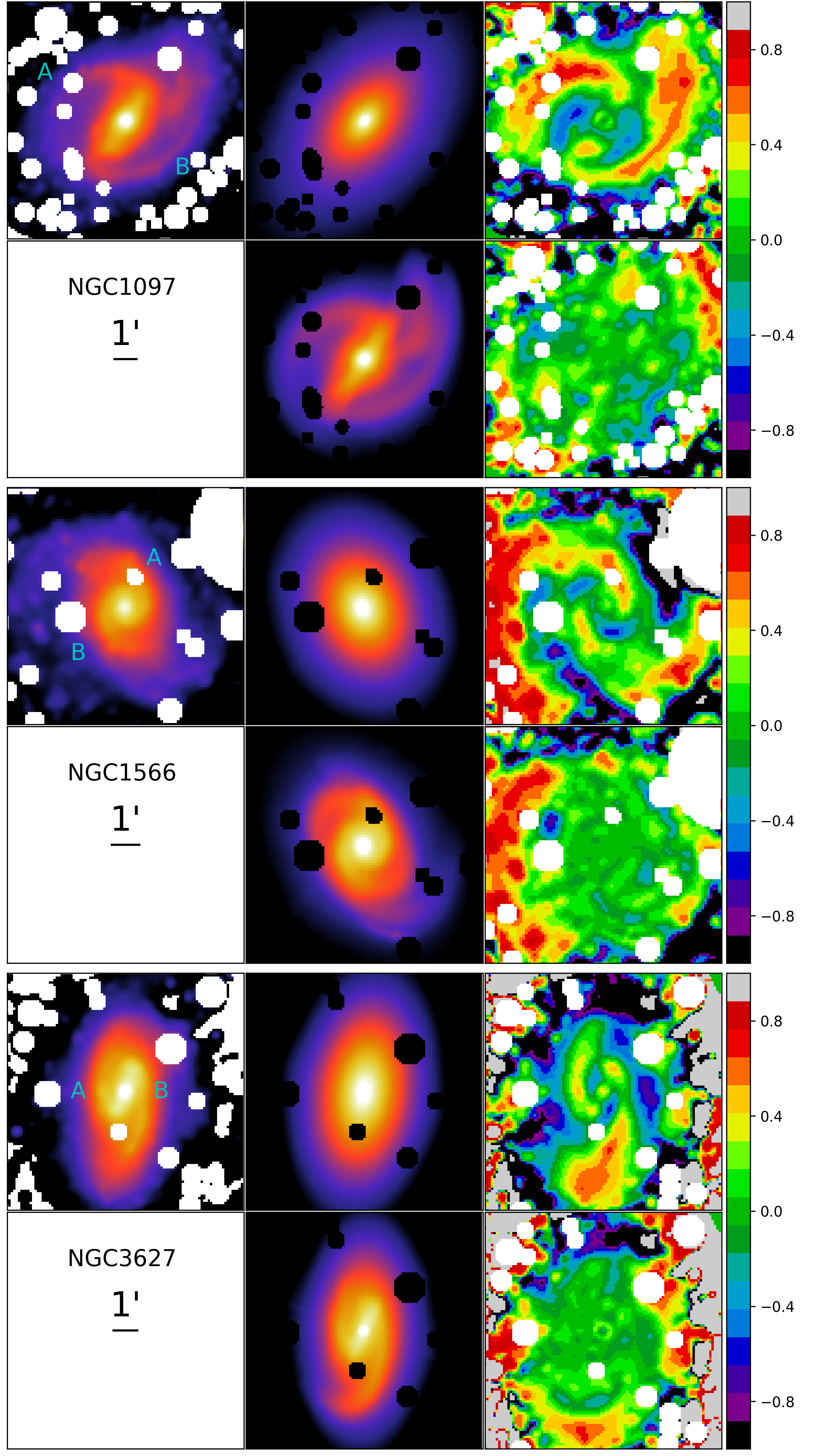}
\caption{Photometric 
decomposition of the galaxies in the 3.6~$\upmu$m band. The~first row shows the original image (uppercase letters A/B mark individual spiral arms), the~S{\'e}rsic-only model, and~the corresponding relative residuals. The~second row presents the full model and its relative residuals. The~relative residual is defined as the difference between the original image and model fluxes, divided by the original image flux. The~color bar on the right indicates the relative residual values, where colors close to green represent good agreement between the model and~observations.\label{fig:2dcolumn}}
\end{figure}

Before performing the decomposition, we applied a series of preparatory steps. The~images in all bands, already convolved with the PSF, were cropped to an appropriate size to improve computational efficiency during the fitting process. For~each image, we manually constructed a mask that covers foreground stars, neighboring galaxies, and~other external objects, as~well as any evident artifacts. In~some cases, we also masked bright star-forming clumpy regions within the galaxy's structure, as~these features cannot be accurately described by a smooth model and often cause the best-fitting spiral arms to deviate from their expected shape (see, for~instance, the~bright spot in Figure
~\ref{fig:2dcolumn} for NGC~3627 in the southern region).
Although it is theoretically possible to fit most of these clumps (see~\cite{2024ApJ...960...25K,2025MNRAS.537..402K}), doing so is computationally challenging and will be considered in future work. During~the decomposition and $\chi^2$ calculation, we utilized the error maps provided in the DustPedia archive, with~a few exceptions where the original sigma maps were directly obtained from the IRSA database\endnote{\url{https://irsa.ipac.caltech.edu/} (accessed on 10 06 2024)}.

We utilized the \verb|IMFIT| package~\citep{Erwin2015} for photometric decomposition in this study. For~each galaxy, we constructed two sets of models: (i) a single-component S{\'e}rsic model, following the approach of~\cite{Mosenkov2019}, and~(ii) a multi-component model incorporating all relevant structural components. These included disks modeled with an exponential function~\citep{Freeman1970}, bulges described by a S{\'e}rsic function~\citep{Sersic1968}, and~bars modeled using a generalized ellipse function, implemented as \verb|Sersic_GenEllipse|. Individual spiral arms were fitted using a specialized function detailed in~\cite{Chugunov2025} and implemented as an \verb|IMFIT| class\endnote{The latest version is available at \url{https://github.com/IVChugunov/IMFIT_spirals} (accessed on 20 03 2025).}.

Additionally, we modeled several of the brightest clumps and the unresolved AGN in NGC~3627~\citep{1995ApJ...443..617S,2008ARA&A..46..475H} as a \verb|PointSource| component, which corresponds to the PSF scaled by a constant factor. The~expected central depression in the galaxy, potentially caused by star formation suppression~\citep{2022A&A...658A..74P} or bar-induced effects, could also be incorporated into the model. However, since exponential disk models provided a satisfactory fit to the data, we did not introduce this~modification. 

The specific steps for fitting spiral arms are described in the next paragraph, while a detailed list of the components used for each galaxy is provided in Table~\ref{tab:heroutput}.

In each of the three galaxies, we modeled two primary spiral arms, with~some specific nuances. In~NGC~1097, we additionally modeled a small segment visible in the upper-right corner. Unlike the other cases, both arms exhibit ``breaks'' in the model, forming a so-called pseudoring around the~bar. 

In NGC~1566, there is a noticeable sharp drop in the brightness of the spiral arms. To~account for this, we modeled the inner and outer sections separately, labeling them as \textit{in} and \textit{out} where necessary. While NGC~1566 does contain a bar, it is relatively small (see, for~example, the~masks in~\cite{2021A&A...656A.133Q}) and is nearly unresolved at the adopted resolution. Consequently, we did not include it in the~model. 

In NGC~3627, the~left spiral arm (Spiral A) is better described by two distinct arms, which become particularly evident at the tips. Therefore, we included an additional faint arm model on the periphery, although~our analysis focuses primarily on the bright inner arm, along with Spiral B. All spiral arms discussed in the following sections are labeled in Figure~\ref{fig:2dcolumn}. 

A crucial aspect of performing decomposition efficiently with such complex functions is selecting appropriate initial values. To~achieve this, we utilized a script that recalculates parameters based on manually identified spiral arms in each image. These identifications rely on the original resolution as well as supplementary data from HST, JWST, and~other sources, including H{\sc i} observations~\citep{2021A&A...656A.133Q,2024A&A...691A.163E}.

The resulting models, presented in Figures~\ref{fig:2dcolumn},~\ref{fig:1097appendix}--\ref{fig:3627appendix}, provide a reasonable fit to the galaxies. Their properties are discussed in detail in Section~\ref{sec:decomp_results}; here, we briefly outline the key aspects that validate these~models.

\textls[-5]{The first validation criterion is the visual appearance of the models. Similar to \mbox{Figure~2} in~\cite{Chugunov2024}, we present the models alongside the original images, along with relative residuals for the 3.6~$\upmu$m band in Figure~\ref{fig:2dcolumn}. As~evident from these residual maps, even in challenging cases, the~residuals remain small, primarily appearing in green. Additionally, the~auxiliary Figures~\ref{fig:1097appendix}--\ref{fig:3627appendix}, which display azimuthal profiles, demonstrate that the model effectively approximates both the individual ``bumps'' and the overall complex structure of the~galaxies.}

The second validation criterion comes from comparing the S{\'e}rsic-only models with previous measurements. The~structural parameters obtained from these models are consistent with those measured in three filters in~\cite{Mosenkov2019}, supporting the reliability of our~pipeline.

Following~\cite{Marchuk_M51}, we analyze the distribution of pixel values in the residual images and find them to be nearly symmetrical, with~median values close to zero---an expected characteristic of well-constrained models. Furthermore, similar to~\cite{Marchuk_M51}, we assess whether the inclusion of a large number of additional parameters is justified. While increasing the model complexity naturally reduces $\chi^2$, we verify its justification using the Bayesian information criterion (BIC,~\cite{1978AnSta...6..461S,2017pbi..book.....B}). The~BIC is computed using a modified formula~\mbox{\citep{2011ApJS..196...11S,2014MNRAS.440.1690H}}:
\begin{equation}
\mathrm{BIC}=\frac{\chi^2}{A_{\mathrm{PSF}}} + k\cdot\ln{\frac{n}{A_{\mathrm{PSF}}}},
\end{equation}
where $A_{\mathrm{PSF}}$ represents the number of pixels in the PSF, $k$ is the number of free parameters, and $n$ is the total number of pixels. In~all cases, the~BIC value is lower for the full model incorporating all components, validating its superiority over the oversimplified S{\'e}rsic-only~version.

Although a full uncertainty estimation via a bootstrap resampling procedure in {\small IMFIT} is computationally expensive, we performed approximately ten iterations. The~results confirm that the estimated parameters remain stable, reinforcing the robustness of the~model.

\begin{figure}[H]
\includegraphics[width=13.8cm]{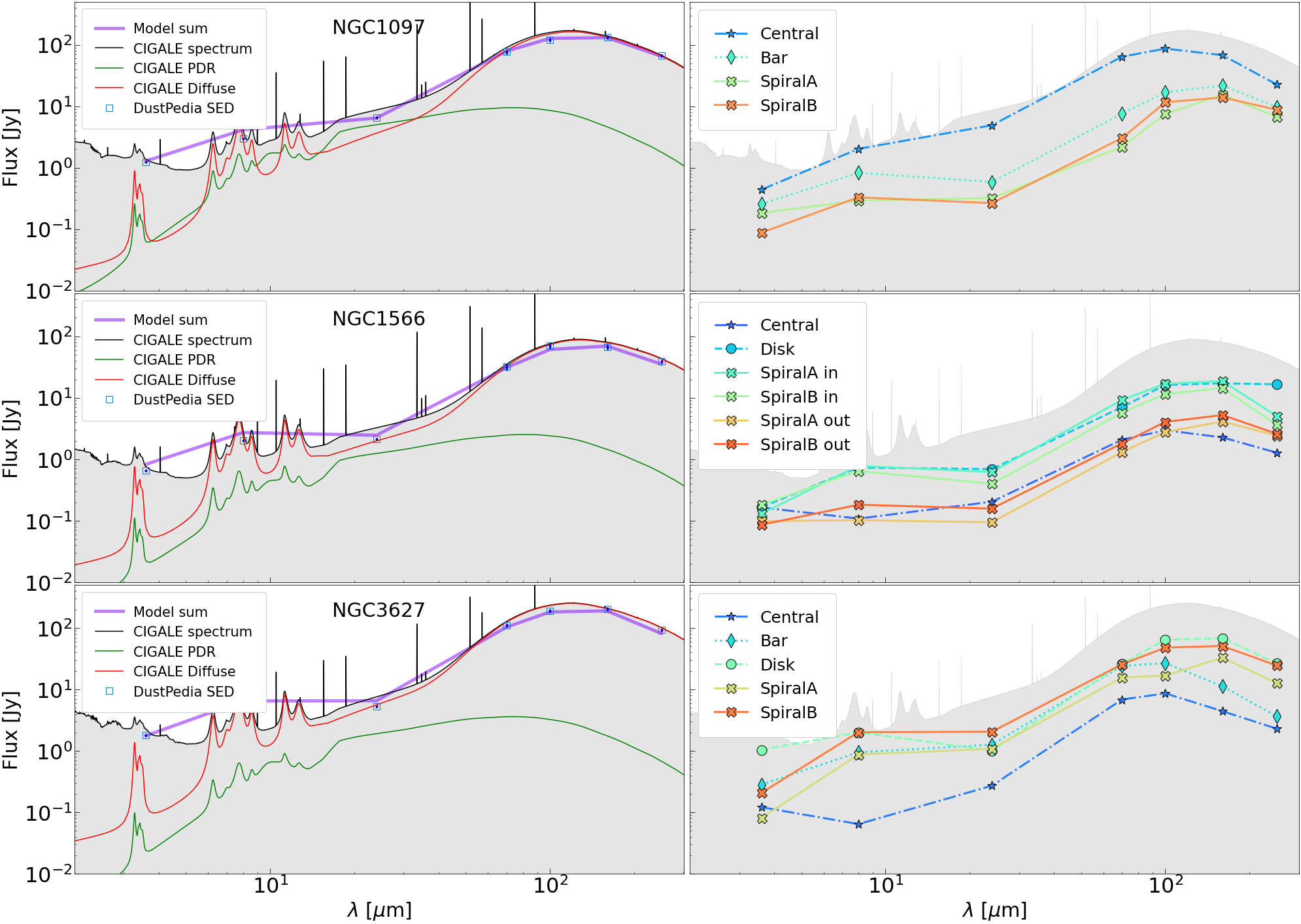}
\caption{SEDs 
for the galaxies and their individual components. The~left column presents the fits obtained by~\cite{2019A&A...624A..80N} using CIGALE with the THEMIS dust model. These fits include the total spectrum and separate contributions from photo-dissociation regions (PDRs) and diffuse dust, as~well as the total flux sum for the components modeled in this work. The~right column displays the total CIGALE fit alongside the SEDs for individual components in each~galaxy. \label{fig:sed}}
\end{figure}

Finally, an~additional key validation comes from Figure~\ref{fig:sed}, which shows that the total \textit{modeled} SED aligns perfectly with the aperture-matched photometry from~\cite{dustpedia}. This agreement indicates that we have accounted for the majority of light sources accurately and~comprehensively.

\subsection{SED Fitting with~HerBIE}
\label{sec:sedfitting}

We performed component-wise dust SED fitting using the \verb|HerBIE| (HiERarchical Bayesian Inference for dust Emission) model~\citep{2018MNRAS.476.1445G}. By~employing a hierarchical Bayesian framework and the Markov chain Monte Carlo (MCMC) technique to infer dust parameters, this approach effectively mitigates certain inherent degeneracies among them. In~previous studies, \verb|HerBIE| has been successfully applied to both global and resolved SED fitting in large samples~\citep{2021A&A...649A..18G,2022A&A...667A..35R}, as~well as to individual objects~\citep{2022EPJWC.25700023K,2023A&A...679A...7K,2024EPJWC.29300038P,2019PASJ...71..123B}.

As in previous studies, the~fitting was performed under the assumption that grain properties follow the \verb|THEMIS| model\endnote{Available at \url{http://www.ias.u-psud.fr/themis/} (accessed on 10 06 2024)}, as~presented in~\cite{2017A&A...602A..46J}. This framework does not intrinsically include polycyclic aromatic hydrocarbons (PAHs); instead, it models analogous small grains, which are partially hydrogenated amorphous carbons, denoted as a-C(:H). Additionally, \verb|THEMIS| includes a population of a-C(:H)-coated large amorphous silicates and large a-C(:H), with~metallic inclusions.  All dust grains are distributed by size according to the SED features that they exhibit in emission (see Figure~1 in~\cite{2021A&A...649A..18G}).

We modeled each component in every galaxy separately using all seven IR bands within the 3.6~$\upmu$m--250~$\upmu$m wavelength range, specifically, three bands from IRAC, three from PACS, and~one from SPIRE. For~each galaxy component, we used the flux estimates obtained during the decomposition process. Since \verb|IMFIT| tends to underestimate uncertainties, we adopted the same relative errors for the individual components as those in the total SED. While this is a simplification, the~obtained results show good agreement with independent sources, suggesting that the impact of this assumption is not critical. Another simplification is that all central components are modeled as the sum of the bulge and all other central structures, including the AGN. As~demonstrated by~\cite{2020A&A...638A.150V,2012MNRAS.420.2756S,2006A&A...452..459H}, accurate radiative transfer modeling of the AGN requires additional assumptions about geometry and physical properties. While this is unlikely to significantly affect our results---since the AGN primarily heats dust outside the torus only in its immediate vicinity (see Figure~7 in~\cite{2020A&A...638A.150V})---all results related to the `Central' components should nonetheless be interpreted with~caution.

The results of the modeling are presented in Table~\ref{tab:heroutput}, and~the example of SED fitting is presented in Appendix~\ref{sec:sedappendix}. For~a detailed definition of the parameters derived by \verb|HerBIE|, we refer the reader to Section~3.1 of~\cite{2021A&A...649A..18G}. Below, we provide a brief description of the key parameters: dust mass ($M_{\mathrm{dust}}$), mean starlight intensity ($\bar{U}$), and~the mass fraction of PAHs~($q_{\mathrm{PAH}}$). 

The dust mass $M_{\mathrm{dust}}$ is estimated under the assumption that it follows a power-law distribution of $U$ and depends on the adopted distance (see Table~\ref{tab:galdata}). The~starlight intensity $U$ quantifies the interstellar radiation field (ISRF) that illuminates the dust grains. It is also related to the equilibrium temperature of large dust grains, assuming a Milky Way-like spectrum, via the relation $T_{\mathrm{dust}} = 18.3[\mathrm{K}] \times U^{(4+\beta)^{-1}}$, where the emissivity index is taken as $\beta=1.7$ \citep{2019A&A...624A..80N}. 
 The \verb|THEMIS| model does not include PAH, but~has a population of small, stochastically heated aromatic-feature-carrying (AF) grains, which demonstrate similar properties. They can be used to estimate PAH mass fraction $q_{\mathrm{PAH}}$ following the relation from~\cite{2021A&A...649A..18G}: $q_{\mathrm{PAH}} = 0.45\times q_{\mathrm{AF}}$. It is important to note that we perform this only in order to compare our results to the literature, and~keep in mind that, in the discussion in Section~\ref{sec:sedresults}, all mentions of $q_{\mathrm{PAH}}$ are actually derived from the amount of small a-C(:H).

\section{Results and~Discussion}
\label{sec:results}

In this section, we present the results of our analysis and discuss them in the context of previous~studies.

\subsection{Decomposition~Results}
\label{sec:decomp_results}

\begin{figure}[H]
\includegraphics[width=13.8cm]{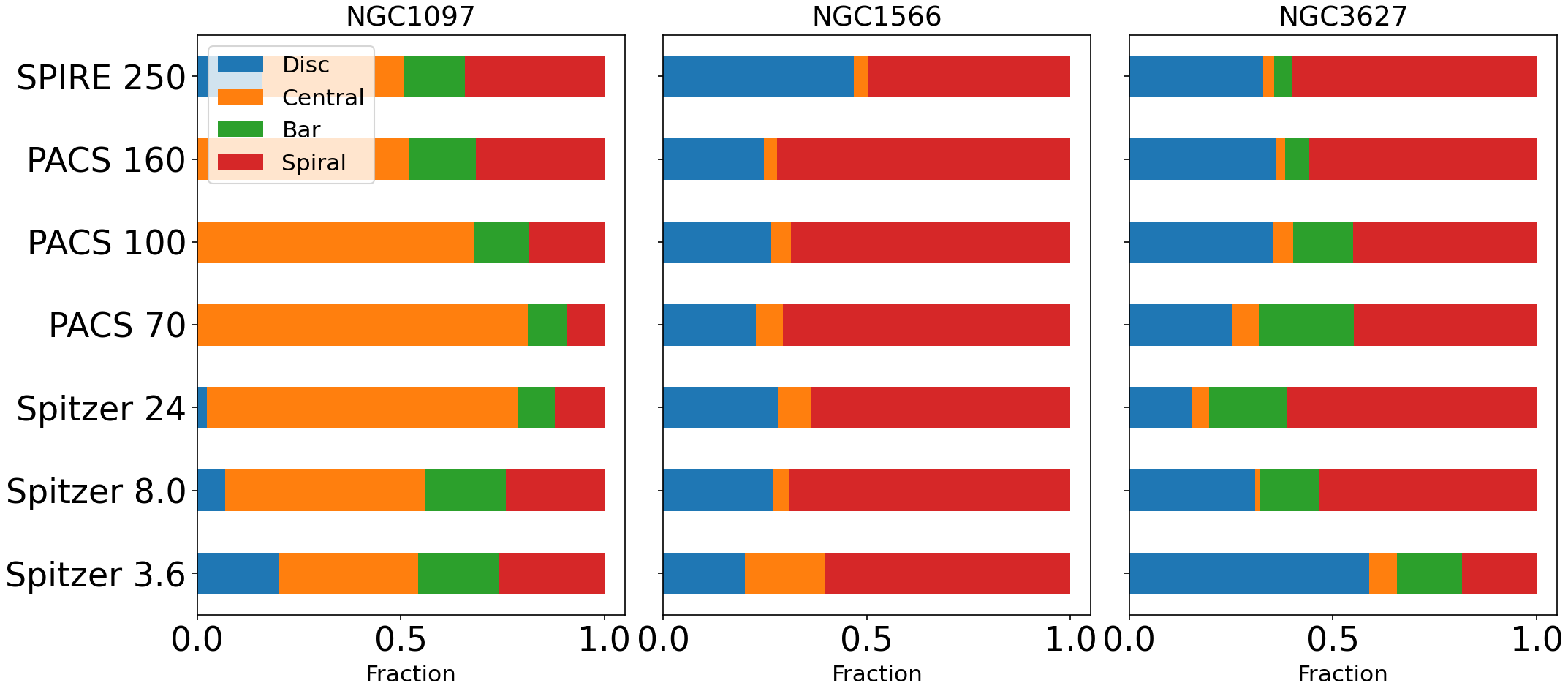}
\caption{Fractions of the total light attributed to individual components, as~measured in the photometric decomposition.
\label{fig:fractions}}
\end{figure}

In Figure~\ref{fig:fractions}, we present the fractional contributions of the different components in the modeled galaxies. Similar to the behavior observed for S{\'e}rsic indices, all three galaxies exhibit distinct structural~characteristics.

For NGC~1097, we identify a highly luminous central region, consisting of a nuclear ring and an AGN, which accounts for more than half of the total light in most IR bands. Another notable feature is that, when spiral arms are included, the~galaxy can be modeled with either a very small disk or, in~some cases, no disk at~all.

In NGC~1566, the~spiral arms emerge as the most luminous component, which is expected given their prominence and extensive structure. In~contrast, the~fractional contribution of the spiral arms in NGC~3627 is lower. However, the~bar and disk components are, on~average, the~most dominant among the~sample.

In the following sections, we examine each galaxy in detail, with~a particular focus on the `dust' IR profile and its deviations from a standard exponential profile~\citep{Mosenkov2019}.

As evident from Figures~\ref{fig:fractions} and~\ref{fig:1097appendix}, NGC~1097 has a significant fraction of its light concentrated in the central region. In~all profiles, we observe a distinct ``bump'', which is typically indicative of an early-type spiral galaxy with a prominent classical bulge~\cite{deVaucouleurs1948}. However, in~the case of NGC~1097, we know that the unresolved inner ring and the active nucleus also contribute significantly to the central brightness. As~shown in Figure~\ref{fig:1097appendix}, this explains why the S{\'e}rsic-only profile exhibits an index of $n > 1$: it primarily captures the central structure and consequently deviates from the profile beyond 150--200~arcsec. In~contrast, the~full model accurately follows the profile across its entire extent, including the outer ``bumps'' produced by the spiral arms, which remain visible even after azimuthal averaging. The~central component is well represented by a S{\'e}rsic profile with $n < 1$.

Another intriguing feature of NGC~1097 is its small or possibly absent dust disk. This arises from the fact that the signal level in the interarm regions, between~the grand-design spiral pattern, is comparable to the noise level outside the galaxy in the same image. This effect is particularly evident in the PACS bands, where significant noise appears in the form of flocculent ``filaments'', similar in structure to those found between the spiral arms. Notably, this characteristic is also present in the images at their original resolution. In~principle, it is possible to model all other images of NGC~1097 without including the exponential disk component. The~resulting models would exhibit only a slight degradation in terms of $\chi^2$ and the~BIC.

For NGC~1566, the~situation is markedly different. The~spiral arms are less pronounced and do not appear as distinct features in the surface brightness profiles, which overall exhibit a shape close to an exponential profile, as~shown in Figure~\ref{fig:1566appendix}. Consistent with this observation, the~S{\'e}rsic-only model yields an index of $n \approx 1$, in~agreement with previous measurements by~\cite{Mosenkov2019} and as illustrated in Figure~\ref{fig:sers}. The~galaxy is well represented by a model consisting of `classical components plus spiral arms'. Another interesting feature is that, in~the azimuthal profile shown in Figure~\ref{fig:1566appendix}, the~spiral arms collectively resemble an additional exponential disk. This effect may be a consequence of their considerable length and the number of full revolutions that they~complete.

Finally, in~the case of NGC~3627, we observe that the S{\'e}rsic index is close to zero in most bands. In~other words, there is a central depression in the dust distribution, similar to what is found in approximately half of the galaxies analyzed in~\cite{Mosenkov2019}. This behavior arises due to a `plateau' extending from the center to the ``bump'' at $\sim70$~arcsec, which forces a lower $n$ value to achieve an optimal fit.
This feature in the profile results from the significant contribution of both spiral arms, making NGC~3627 an example where the presence of spiral arms leads IR profiles to favor a Gaussian function over an exponential one. In~fact, the~possibility of such a profile for cold dust emission was previously discussed in Section~5.4.1 of~\cite{Mosenkov2019} and in the case of M~81 by~\cite{2010A&A...518L..64S}.

Before proceeding further, we highlight several similarities with previous studies. According to Figure~\ref{fig:fractions}, the~spiral-to-total fraction ($S/T$) constitutes approximately 20--60\% of the total light, which is consistent with the estimates from~\cite{Marchuk_M51,Chugunov2024,Savchenko2020}. Moreover,~\cite{Chugunov2024} found (see Figure~8 in their work) that, in the 3.6~$\upmu$m Spitzer band, the~$S/T$ ratio reaches its maximum for galaxies with Hubble types $T \approx 3$--5. This trend is also observed in our study, except~when considering the central components of NGC~1097.

The parameters for the spiral arms are listed in Table~\ref{tab:spiralparams}. Similar to the multiwavelength analysis of M~51 conducted by~\cite{Marchuk_M51}, we find that the width of the spiral arms is smallest in the 24--70~$\upmu$m range. This can be connected with the fact that spiral arms in these bands are associated with ongoing star
formation, which cannot propagate far enough from the place of birth~\cite{2023A&A...673A.147P}.  Additionally, the~pitch angle remains roughly constant across wavelengths, though~its variation along the arm is significant (see also~\cite{Savchenko2013}). This aspect is further discussed in Section~\ref{sec:cr}.


\begin{table}[H] 
\caption{Measured parameters for the models of individual spiral~arms. \label{tab:spiralparams}}
\begin{tabularx}{\textwidth}{lCCCC}
\toprule
\textbf{Galaxy/Spiral} & \textbf{Band} &  \textbf{Spiral/Total} &  \textbf{Pitch \boldmath$\psi$, Deg} &  \textbf{Width, Arcsec}\\
\midrule
NGC~~1097 A & Spitzer 3.6 & 0.14 & $17.88\pm1.82$ & $69.92\pm0.69$ \\
NGC~~1097 A & Spitzer 8.0 & 0.07 & $13.89\pm1.06$ & $24.07\pm1.43$ \\
NGC~~1097 A & Spitzer 24 & 0.05 & $14.11\pm1.07$ & $30.27\pm3.15$ \\
NGC~~1097 A & PACS 70 & 0.03 & $13.58\pm1.24$ & $20.18\pm0.32$ \\
NGC~~1097 A & PACS 100 & 0.06 & $12.70\pm1.42$ & $27.91\pm0.33$ \\
NGC~~1097 A & PACS 160 & 0.12 & $14.37\pm1.21$ & $33.83\pm0.44$ \\
NGC~~1097 A & SPIRE 250 & 0.10 & $13.59\pm1.10$ & $22.26\pm0.85$ \\
NGC~~1097 B & Spitzer 3.6 & 0.07 & $18.25\pm1.67$ & $50.86\pm0.68$ \\
NGC~~1097 B & Spitzer 8.0 & 0.08 & $20.51\pm0.45$ & $33.33\pm1.08$ \\
NGC~~1097 B & Spitzer 24 & 0.04 & $21.34\pm0.32$ & $51.25\pm4.72$ \\
NGC~~1097 B & PACS 70 & 0.04 & $17.70\pm0.58$ & $51.78\pm0.00$ \\
NGC~~1097 B & PACS 100 & 0.09 & $22.05\pm1.75$ & $58.90\pm0.00$ \\
NGC~~1097 B & PACS 160 & 0.11 & $20.94\pm0.43$ & $45.92\pm0.49$ \\
NGC~~1097 B & SPIRE 250 & 0.13 & $18.13\pm0.78$ & $31.47\pm0.78$ \\
\midrule
NGC~1566 A$_{in}$ & Spitzer 3.6 & 0.16 & $27.79\pm2.70$ & $47.03\pm1.30$ \\
NGC~1566 A$_{in}$ & Spitzer 8.0 & 0.29 & $19.93\pm5.96$ & $34.21\pm0.65$ \\
NGC~1566 A$_{in}$ & Spitzer 24 & 0.25 & $30.17\pm3.92$ & $39.12\pm2.36$ \\
NGC~1566 A$_{in}$ & PACS 70 & 0.30 & $24.75\pm4.84$ & $29.82\pm0.00$ \\
NGC~1566 A$_{in}$ & PACS 100 & 0.28 & $24.75\pm4.84$ & $33.22\pm0.00$ \\
NGC~1566 A$_{in}$ & PACS 160 & 0.27 & $24.75\pm4.84$ & $39.80\pm0.63$ \\
NGC~1566 A$_{in}$ & SPIRE 250 & 0.14 & $21.99\pm3.92$ & $14.87\pm2.37$ \\
NGC~1566 B$_{in}$ & Spitzer 3.6 & 0.22 & $13.97\pm7.34$ & $58.32\pm0.00$ \\
NGC~1566 B$_{in}$ & Spitzer 8.0 & 0.24 & $11.16\pm15.63$ & $42.22\pm0.97$ \\
NGC~1566 B$_{in}$ & Spitzer 24 & 0.16 & $29.56\pm2.89$ & $34.62\pm2.96$ \\
NGC~1566 B$_{in}$ & PACS 70 & 0.18 & $26.80\pm5.13$ & $30.34\pm0.00$ \\
NGC~1566 B$_{in}$ & PACS 100 & 0.19 & $26.80\pm5.13$ & $33.79\pm0.00$ \\
NGC~1566 B$_{in}$ & PACS 160 & 0.21 & $26.80\pm5.13$ & $42.95\pm0.00$ \\
\bottomrule
\end{tabularx}
\end{table}

\begin{table}[H]\ContinuedFloat
\caption{{\em Cont.}\label{tab:spiralparams}}
\begin{tabularx}{\textwidth}{lCCCC}
\toprule
\textbf{Galaxy/Spiral} & \textbf{Band} &  \textbf{Spiral/Total} &  \textbf{Pitch \boldmath$\psi$, Deg} &  \textbf{Width, Arcsec}\\
\midrule
NGC~1566 B$_{in}$ & SPIRE 250 & 0.10 & $25.13\pm2.64$ & $12.16\pm2.26$ \\
NGC~1566 A$_{out}$ & Spitzer 3.6 & 0.12 & $14.34\pm0.50$ & $58.41\pm0.00$ \\
NGC~1566 A$_{out}$ & Spitzer 8.0 & 0.04 & $18.54\pm0.60$ & $36.21\pm2.98$ \\
NGC~1566 A$_{out}$ & Spitzer 24 & 0.04 & $17.68\pm0.38$ & $38.39\pm0.00$ \\
NGC~1566 A$_{out}$ & PACS 70 & 0.04 & $17.23\pm0.32$ & $28.35\pm0.00$ \\
NGC~1566 A$_{out}$ & PACS 100 & 0.05 & $17.23\pm0.32$ & $31.58\pm0.00$ \\
NGC~1566 A$_{out}$ & PACS 160 & 0.06 & $17.23\pm0.32$ & $40.14\pm0.00$ \\
NGC~1566 A$_{out}$ & SPIRE 250 & 0.07 & $18.54\pm0.86$ & $44.39\pm6.31$ \\
NGC~1566 B$_{out}$ & Spitzer 3.6 & 0.10 & $10.88\pm0.00$ & $57.63\pm0.00$ \\
NGC~1566 B$_{out}$ & Spitzer 8.0 & 0.07 & $12.15\pm0.00$ & $42.61\pm0.00$ \\
NGC~1566 B$_{out}$ & Spitzer 24 & 0.06 & $23.69\pm0.00$ & $39.94\pm0.00$ \\
NGC~1566 B$_{out}$ & PACS 70 & 0.06 & $11.89\pm0.00$ & $27.67\pm0.00$ \\
NGC~1566 B$_{out}$ & PACS 100 & 0.07 & $11.89\pm0.00$ & $30.83\pm0.00$ \\
NGC~1566 B$_{out}$ & PACS 160 & 0.08 & $11.89\pm0.00$ & $39.18\pm0.00$ \\
NGC~1566 B$_{out}$ & SPIRE 250 & 0.07 & $14.80\pm0.00$ & $37.56\pm4.68$ \\
\midrule
NGC~3627 A & Spitzer 3.6 & 0.07 & $18.23\pm1.54$ & $54.96\pm0.00$ \\
NGC~3627 A & Spitzer 8.0 & 0.22 & $17.17\pm0.94$ & $16.46\pm0.02$ \\
NGC~3627 A & Spitzer 24 & 0.30 & $26.26\pm0.50$ & $23.39\pm11.56$ \\
NGC~3627 A & PACS 70 & 0.20 & $27.09\pm0.17$ & $13.08\pm0.45$ \\
NGC~3627 A & PACS 100 & 0.19 & $19.57\pm0.30$ & $16.15\pm1.21$ \\
NGC~3627 A & PACS 160 & 0.29 & $14.14\pm1.73$ & $29.91\pm0.71$ \\
NGC~3627 A & SPIRE 250 & 0.30 & $11.76\pm2.20$ & $25.06\pm1.36$ \\
NGC~3627 B & Spitzer 3.6 & 0.12 & $22.35\pm0.25$ & $56.44\pm0.00$ \\
NGC~3627 B & Spitzer 8.0 & 0.31 & $13.75\pm0.30$ & $29.02\pm0.01$ \\
NGC~3627 B & Spitzer 24 & 0.31 & $17.45\pm0.06$ & $23.87\pm3.79$ \\
NGC~3627 B & PACS 70 & 0.25 & $16.83\pm0.05$ & $23.15\pm0.23$ \\
NGC~3627 B & PACS 100 & 0.26 & $15.01\pm0.22$ & $26.28\pm0.39$ \\
NGC~3627 B & PACS 160 & 0.27 & $20.30\pm0.18$ & $29.00\pm0.41$ \\
NGC~3627 B & SPIRE 250 & 0.30 & $18.55\pm0.26$ & $24.60\pm0.78$ \\
\bottomrule
\end{tabularx}
\end{table}

\subsection{Presence of Density~Wave}
\label{sec:cr}

Measurements of the properties of spiral arms can provide insights into the nature of spiral structure in galaxies. The~general concept is that, if a galaxy exhibits a long-lived density wave (DW) rotating with a constant pattern angular velocity, $\Omega_p$, then the pattern rotates faster than the disk at large distances from the center and more slowly than the disk in the inner regions. 
This behavior produces an observable effect: beyond the so-called corotation radius (CR), where the disk and spiral pattern angular velocities are equal, newly formed stars may appear to move downstream relative to the main population of older stars. If~the galaxy's rotation curve is known, estimating the CR is equivalent to determining the corresponding pattern speed, $\Omega_p$.

Based on this fundamental idea, various observational methods have been employed to determine the CR. These include measuring the skewness of the intensity profile along the spiral arm~\citep{Marchuk2024}, analyzing color gradients along the arm~\citep{2009ApJ...707.1650M}, and~identifying offsets between young star clusters and their nearest H~\textsc{ii} regions~\citep{2021MNRAS.508..912S}, among~other techniques. However, the~most common approach involves comparing the pitch angle, $\psi$, in~different wavelength bands.
The general consensus is that the pitch angle in redder bands, which trace the older stellar population, should be larger than in bluer bands, which are associated with ongoing star formation~\citep{2018ApJ...869...29Y}. However, it is important to note that this simplified picture can be more complex due to additional physical processes~\citep{Miller2019,2023MNRAS.524...18M}.


As noted by~\cite{Savchenko2013}, spiral arms can adopt various shapes~\citep{1981AJ.....86.1847K} (see also Chugunov~et~al., in~prep.). Therefore, it is important to emphasize that our measurements account for the total distribution of light across the spiral arms. As~a tracer of the old stellar population, we use the Spitzer 3.6~$\upmu$m band. For~indicators associated with star formation, we use MIR images in the 8~$\upmu$m, 24~$\upmu$m, and~70~$\upmu$m bands, where warm dust is believed to be heated by UV photons from young stars~\citep{2007ApJ...666..870C,Tamburro2008}. The~total FIR emission is also commonly used as such an indicator~\citep{2009ApJ...703.1672K,1990ApJ...350L..25D}. However, since we do not analyze wavelengths beyond 250~$\upmu$m due to low resolution, we do not include it in our~study.

According to the data in Table~\ref{tab:spiralparams}, we find no galaxy in which the pitch angle, $\psi$, at~3.6~$\upmu$m is consistently larger than in any of the star-formation-related indicators for both arms. Given the substantial variation in the pitch angle, we also estimate it over the entire extent of each arm and arrive at the same~conclusion.

This result can be interpreted as evidence against the presence of a stationary density wave in these galaxies. Additional arguments supporting this interpretation, along with comparisons to previous studies, are presented below. However, before~addressing these points, we first assess whether this effect can be detected in principle, given the resolution limitations of our data.
To investigate this, we follow the approach of~\cite{Marchuk_M51}. Assuming a flat rotation curve, $v(R) = v_0$, a~pitch angle $\psi$ in one indicator, and~a time delay $\Delta t$ corresponding to the time a star takes to cross the arm and evolve~\citep{Tamburro2008}, we can determine the modified pitch angle, $\psi\prime$, for~a second indicator using the following formula:
$$\psi\prime = \mathrm{arctan}\left(\frac{1}{\tan \psi} \pm \frac{\Omega\times(1-\beta)}{\log \beta}\times\Delta t \right)^{-1}\,$$
where $\beta=R_{out}/R_{CR}$ is the ratio of the arm length to the corotation distance, and \mbox{$\Omega=v_0/R_{out}$}. The~$\pm$ sign indicates that the azimuthal offset can occur in either direction. For~the plateau values of the rotation curve, $v_0$, we adopt 250~km/s for NGC~1097~\cite{2003ApJ...585..281H}, 180~km/s for NGC~1566~\cite{2019MNRAS.487.2797E}, and~200~km/s for NGC~3627~\cite{2008AJ....136.2648D}. Using the pitch angles and arm extensions from Table~\ref{tab:spiralparams}, assuming that the CR is located at the midpoint of the arm and using $\Delta t = 5/10/20/50/100$~Myr, we estimate the difference $\Delta \psi = |\psi - \psi\prime|$. We find that, for NGC~1566 and NGC~3627, the~average $\Delta \psi \approx 5^{\circ}$ for $\Delta t = 10$~Myr, which should be detectable for all time intervals equal to or greater than this. For~NGC~1097, we measure $\Delta \psi \approx 3^{\circ}$ for $\Delta t = 20$~Myr. While the effect in this galaxy is smaller, the~well-constrained shape of the spiral arms suggests that this offset should still be~detectable.

Another related method, which is in fact more direct and can also be applied in this study, involves measuring the precise angular offsets between the same two indicators. As~demonstrated by~\cite{Tamburro2008,Foyle2011}, in~the case of a density wave, these offsets should follow the following:
\begin{linenomath}
\begin{equation}
   \Delta \phi(R) = (\Omega(R)-\Omega_{\rm p}) \times \Delta t,
    \label{eq:offset}
\end{equation}
\end{linenomath}
where $\Omega(R)$ and $\Omega_{\rm p}$ are the angular velocities of the disk and the spiral pattern, respectively, and~$\Delta t$ represents the time interval between two evolutionary stages of the stellar population. If~the rotation curve is known---which is available for all three galaxies from~\mbox{\cite{2003ApJ...585..281H,2019MNRAS.487.2797E,2008AJ....136.2648D}}---we can derive the exact form of the angular offset profile and estimate the location of the CR at the point where the offset crosses zero, $\Delta \phi(R) = 0$. This method has been shown to be effective and generally consistent with other approaches (see~\cite{Kostiuk2025}), though~it has certain limitations \citep{Foyle2011}. However, these shortcomings can be mitigated by incorporating the full light distribution for proper modeling. We measure these angular offsets using the same indicators as those employed for the pitch angle analysis. For~nearly all spiral arms, we observe a consistent pattern: the offset is oriented in one direction, with~the 3.6~$\upmu$m band tracing the outer convex part of the arm, while the star-formation-related images occupy the inner concave region. The~only exceptions are Spiral $A_{out}$ in NGC~1566, where we detect a zero-crossing that does not align with the expected rotation curve, and~Spiral $A_{in}$ in NGC~1566, where the indicators appear to switch positions. However, in~this case, the~offsets are too small to be reliably distinguished. This observed behavior is consistent with certain tidal interaction scenarios, where kinematic spiral arms rotate with a lagging pattern speed, $\Omega_p = \Omega(R) - \varkappa/2$, where $\varkappa$ is the epicyclic frequency~(\citep{Oh2008} and references therein). 
Under~this model, we expect to observe similar pitch angles and unidirectional offsets, as~seen in our~measurements.

Additionally, we highlight other arguments that support our results. Ref.~\cite{Kostiuk2024} found that measurements of the CR estimated using different methods often show significant discrepancies. In~Table~\ref{tab:crsparams}, we list various CR measurements collected from the literature. Most of these data points were obtained using the only direct method proposed by Tremaine and Weinberg~\citep{1984ApJ...282L...5T}.

\begin{table}[H] 
\caption{Corotation parameters from~\cite{Kostiuk2024} and other sources. The~method acronyms are as follows: `P-D' refers to the azimuthal profile phase method; `T-W' represents the Tremaine--Weinberg method~\citep{1984ApJ...282L...5T}; `offset' corresponds to methods similar to those discussed in Section~\ref{sec:cr}; `morph' denotes symmetry-related methods;  `F-B' refers to the velocity phase reversal method~\citep{Font2014a}; `model' includes various modeling approaches; `torque' estimates gravitational torques exerted by the potential on the gas~\citep{2005A&A...441.1011G}. Further details can be found in the referenced papers. Note that some CR measurements apply to the bar and are included in cases where the spiral pattern exhibits the same pattern~speed. \label{tab:crsparams}}
\begin{tabularx}{\textwidth}{LLCCC}
\toprule
\textbf{Galaxy}  & \textbf{\#} & \textbf{CR, Arcsec} & \textbf{Method} & \textbf{Ref.}\\ 
\midrule
NGC~~1097 & 1 & $96.60\pm30.5$ &  T-W     &  \cite{2021AJ....161..185W}         \\
\ldots & 2 & $234.3\pm8.1$ & morph & \cite{1995ApJ...445..591E} \\
\ldots & 3 & $142.8\pm0.0$ & model & \cite{2013ApJ...771....8L} \\
\ldots & 4 & $114.5\pm7.1$ & F-B   & \cite{2014MNRAS.444L..85F} \\
\ldots & 5 & $122.7\pm7.1$ & model & \cite{2014MNRAS.438..971P} \\
\ldots & 6 & $97.5\pm12.2$ & torque & \cite{PHANGSresonances2024} \\
\midrule
NGC~1566 & 1 & $44.40 \pm 26.64$ &  T-W     &  \cite{2021AJ....161..185W}         \\
\ldots & 2 & $78.78 \pm 31.8$ &  T-W     &  \cite{2021AJ....161..185W}        \\
\ldots & 3 & $46.98 \pm 29.22$ &  T-W     &  \cite{2021AJ....161..185W}        \\
\ldots & 4 & $98.70 \pm 5.1$ &  morph   &  \cite{1995ApJ...445..591E}  \\
\ldots & 5 & $97.20 \pm 4.1$ &  P-D     &  \cite{2001ApJ...547..187V}  \\
\ldots & 6 & $122.17 \pm 45.45$ &  offset  &  \cite{2020MNRAS.496.1610A}        \\
\ldots & 7 & $127.39 \pm 36.38$ &  offset  &  \cite{2020MNRAS.496.1610A}           \\
\ldots & 8 & $72.57 \pm 7.0$ &  torque  &  \cite{PHANGSresonances2024}           \\
\midrule
NGC~3627 & 1 & $34.64 \pm 1.46$ &  T-W     &  \cite{2021AJ....161..185W}        \\
\ldots & 2 & $77.08 \pm 26.34$ &  T-W     &  \cite{2021AJ....161..185W}         \\
\ldots & 3 & $85.86 \pm 8.78$ &  T-W     &  \cite{2021AJ....161..185W}    \\
\ldots & 4 & $94.64 \pm 55.62$ &  T-W     &  \cite{2021AJ....161..185W}       \\
\ldots & 5 & $171.00 \pm 28.5$ &  offset  &  \cite{Tamburro2008}                       \\
\ldots & 6 & $163.00 \pm 0.0$ &  T-W     &  \cite{2004ApJ...614..142R}          \\
\ldots & 7 & $95.0 \pm 3.7$ &  torque  &  \cite{PHANGSresonances2024}           \\
\bottomrule
\end{tabularx}
\end{table}

For NGC~1097, we only have two CR measurements; however, a~similar situation is observed across all galaxies: the estimated corotation radii are scattered throughout the disk without a clear consensus, and~their uncertainties encompass either the entire disk or a significant fraction of it (see images in the~\cite{Kostiuk2024} repository). 

If we assume a single CR---which is not necessarily the case~\citep{Marchuk_resonances}---such disagreement may indicate the dynamic nature of the spiral arms (as suggested for NGC~4321 in~\cite{Kostiuk2025}). Alternatively, other explanations may exist, but~the observed discrepancies do not support the presence of a well-defined and strong stationary density~wave.

Ref.~\cite{2000ApJ...541..565K} investigated global spiral modes in NGC~1566 and found that the disk is stable against spiral perturbations under a linear stability analysis~framework. 

Additionally, we performed a simple test using the manual slicing technique described in~\cite{Savchenko2020}. The~objective was to fact-check the position of the spiral arms using a more traditional star formation tracer, namely the GALEX FUV image, which could potentially reveal a clearer offset signal. We measured similar pitch angles in the GALEX FUV band as those obtained previously, as~well as in the Spitzer 3.6~$\upmu$m band, where we also applied the slicing technique. Across all seven measured arms (with only one studied in NGC~3627), we found $\psi_{3.6} \leq \psi_{FUV}$, which we interpret as further confirmation of our results.
The study in \cite{2018ApJ...869...29Y} measured pitch angles for approximately 80 galaxies, including NGC~1097 and NGC~1566, using both one-dimensional and two-dimensional Fourier transformation methods. Their analysis covered a wide range of wavebands, namely 3.6~$\upmu$m, IRVB, NUV, and~FUV, detecting a mild decrease in pitch angle from the reddest to the bluest bands in general. However, in~the two galaxies examined here, no such trend was observed (see Table~1 in~\cite{2018ApJ...869...29Y}), except~in the 1D method for NGC~1097. In~this case, a~very narrow radial range (120--181~arcsec) was used, making the result less~reliable.

Finally, all three galaxies reside in galaxy groups and exhibit signs of tidal interactions. Collectively, this evidence {may} suggest a tidal origin for the spiral patterns in NGC~1097, NGC~1566, and~NGC~3627.

\subsection{SED Fitting~Results}
\label{sec:sedresults}

The dust properties of individual components are presented in Table~\ref{tab:heroutput}. Before~proceeding with further discussion, we note that resolved dust SED fitting has been performed in previous studies~\citep{2022ApJ...926...81A,2024A&A...687A.255S}, though~typically on relatively small samples. One exception is the recent work by~\cite{2025ApJS..276....2C}, in~which the authors estimated the SED for approximately 800 nearby galaxies, including NGC~3627 and NGC~1566, over~the same wavelength range of 3.6~$\upmu$m--250~$\upmu$m. However, even when a resolved SED is available for a given object, the~component-wise fitting approach applied in this study remains valuable due to several advantages:  (i) the number of fitted distributions is significantly smaller, making it easier to control the quality of the fit;  
(ii) to derive information about the properties of individual components, they must still be separated from one~another.

All three galaxies analyzed in this study are late-type systems with $T \approx 3$--4 and thus contain significant amounts of dust and gas, placing them near the peak of the distribution among Hubble stages (see Figure~6 in~\cite{2019A&A...624A..80N}). As~evident from Table~\ref{tab:heroutput}, the~largest fraction of large-grain dust is contained within the disk. However, the~spiral arms also contribute a substantial portion of the total dust mass, $M_{dust}$, with~each arm containing approximately $10^6$--$10^{6.5} M_{\odot}$ ($\sim$10\% of the total dust mass in NGC~1097 and NGC~1566, and~$\sim$20\% in NGC~3627). This dust is distributed over a more compact area, leading to enhanced spiral arm contrast (see, e.g.,~\cite{radiativeM51,2024EPJWC.29300038P,2024A&A...692A..39V}). 
The total dust mass, $M_{dust}$, obtained by summing the contributions from individual components is in good agreement with the estimate from~\cite{2019A&A...624A..80N} for NGC~1566. However, for~NGC~1097, our estimate is systematically higher, whereas, for NGC~3627, it is lower by the same 0.3~dex. All values have been corrected for distance.
 This discrepancy may arise from uncertainties, residual unaccounted light, differences in background subtraction, or~the so-called ``Matryoshka effect'', 
 which highlights the impact of spatial resolution on constrained dust parameters~\citep{2018MNRAS.476.1445G}. Nevertheless, the~discrepancy between the $M_{dust}$ estimates from~\cite{2019A&A...624A..80N,2025ApJS..276....2C} is even larger, suggesting that our results remain within the expected accuracy of SED fitting. Furthermore, the~estimates from~\cite{2020MNRAS.496.3668D} for NGC~1097 and NGC~3627 are in close agreement with those from~\cite{2019A&A...624A..80N}.

Regarding PAHs, we note several key findings. The first and most evident result is that $q_{\mathrm{PAH}}$ is lower in the central regions, as~indicated by SED fitting. This behavior is expected, as~PAH formation via shattering is inefficient in these environments, while depletion through coagulation is significant~\citep{2024A&A...689A..79M,2024EPJWC.29300038P}.  

The second finding is somewhat surprising: in all three galaxies, the~average fraction of these small grains is similar in both the disk and the spiral arms. This result is confirmed by the maps from~\cite{2025ApJS..276....2C}, where the spiral arms exhibit little contrast, except~in the densest star-forming regions, where $q_{\mathrm{PAH}}$ is lower. Furthermore, these maps reveal a gradual increase in the PAH fraction toward the periphery of the disk, reaching approximately 8\%. Similar results were obtained for NGC~3627 by~\cite{2022ApJ...926...81A}. Indeed, PAH emission remains strong in interarm regions, contributing up to 90\% of the total MIR emission, as~observed in M\,83~\citep{2005A&A...441..491V}. PAHs are expected to be efficiently destroyed in H\textsc{ii} regions, which are abundant in spiral arms~\cite{2023ApJ...944L..16E}. However, it is also possible that the total IR emission is not dominated by H~\textsc{ii} regions, as~photodissociation regions (PDRs) are also numerous in spiral arms. The~SED of H~\textsc{ii} regions can, to~some extent, be modeled with a hot modified black-body (MBB) model~\citep{2019A&A...624A..80N} at approximately 30~K. These regions contribute to the MIR continuum between 20~$\upmu$m and 60~$\upmu$m but,~overall, the~dust mass from the neutral medium (constrained in the fitting by far-IR emission) and the PAH fraction (controlled by the 8~$\upmu$m window) are not significantly affected by the presence of H~\textsc{ii} regions. For~the inner bright regions of the spiral arms in NGC~1566, we observe a noticeable increase in $q_{\mathrm{PAH}}$. However, the~arm-averaged value remains close to that of the disk. Additionally, the~finding that $q_{\mathrm{PAH}}$ is similar between the disk and spiral arms in all three galaxies may be due to the fact that, on~large scales, PAH abundance is primarily regulated by metallicity. According to~\cite{2018ARA&A..56..673G}, at~solar metallicity, we indeed expect $\ln(q_\mathrm{AF})\approx-1.7$. 

In NGC~3627, the~bar is visibly deficient in PAHs, likely due to strong dust heating at its tips, which is also evident in the maps from~\cite{2025ApJS..276....2C}. This feature is common in barred galaxies~\citep{2020A&A...637A..25N}.

The dust temperature, evaluated from the interstellar radiation field (ISRF) parameter $\bar{U}$, indicates that dust in the spiral arms is relatively cold, with~$T_{dust} \approx 18$--20~K. This trend is also evident from the SEDs in Figure~\ref{fig:sed}, where the peak of FIR emission for each spiral arm is shifted to longer wavelengths compared to the total fit. Additionally, Figure~\ref{fig:fractions} demonstrates that the fractional contribution of spiral arms increases in bands where dust is~colder.

For NGC~1566, we observe a temperature gradient, with~an inner dust temperature of approximately 25~K decreasing by $\sim$5~K toward the periphery. In~NGC~3627, the~overall decrease in $T_{dust}$ from the center outward is consistent with the spatially resolved SED fitting results reported by~\cite{2022ApJ...926...81A} (see their Figure~10). Furthermore, based on the fits from~\cite{2019A&A...624A..80N}, it is evident that, in each case, the~dust contained within the spiral arms is systematically colder than the average dust temperature in the~galaxy.

The connection between dust properties and the nature of spiral patterns remains poorly understood. Only a few models that explore this relationship exist~\citep{2006NewA...11..240V,Scarano2013}, primarily linking dust properties to the corotation radius and the presence of dust lanes on one side of the spiral arms. In~our own galaxy, observations indicate that the spiral pattern sorts dust by size and fragments it in such a way that different types of grains accumulate on opposite edges of the spiral arms~\citep{2013AstL...39...83G}. 

Other studies suggest that the distribution of supernovae, which play a role in regulating dust content, may also be influenced by the spiral pattern~\citep{2016MNRAS.459.3130A}. This is particularly relevant for NGC~1566, which has hosted an exceptionally high number of supernovae, though~the full implications of this remain under debate. Another avenue of investigation is whether star formation in spiral arms is enhanced due to the velocity gradient induced by the density wave, which could be examined by measuring the arm-to-interarm contrast~\citep{2024A&A...687A.293Q}. 

The dust-to-gas mass ratio (D/G) can be estimated from metallicity. For~these three galaxies, 12 + log(O/H) $\approx 8.55$~\citep{2019A&A...624A..80N}, implying a D/G ratio of several $\times 10^{-3}$~\citep{2021A&A...649A..18G}. This value could be compared with the star formation rate contrast to determine whether star formation is enhanced in the spiral arms~\cite{2010ApJ...725..534F}. In~principle, this hypothesis could be tested in future studies with a larger sample of~galaxies.

\begin{table}[H] 
\caption{Dust parameters of individual components derived from HerBIE SED fitting. A~description of the individual parameters is provided in the last paragraph of Section~\ref{sec:sedfitting}. All values are expressed in natural logarithm, except~for the total dust mass, which is given in decimal logarithm, with~$\Delta$ representing the uncertainty. For~the total values of each galaxy, the~temperature data are taken from~\cite{2019A&A...624A..80N}, while other parameters are from~\cite{2021A&A...649A..18G}. {For comparison, the~last column shows sum of dust mass estimates in this work (decimal logarithm) and fraction of each component.} Dust masses have been corrected for the distances listed in Table~\ref{tab:galdata}. \label{tab:heroutput}}
\begin{tabularx}{\textwidth}{lCCCCCCC}
\toprule
\textbf{Component} & \boldmath$\mathrm{ln} q_{\mathrm{AF}}$ & \boldmath$\Delta$ & $T_{\mathrm{dust}}$ & \boldmath$\Delta$ & \boldmath$\log M_{\mathrm{dust}}$ & \boldmath$\Delta$ & \textbf{Total}\\
\midrule
NGC1097 & $-$2.33 & 0.02 & 24.47 & 0.69 & 7.71 & 0.02 & 7.42  \\
\midrule
Central & $-$2.52 & 0.11 & 27.63 & 1.68 & 6.51 & 0.15 & 12\% \\
Bar & $-$1.65 & 0.09 & 21.42 & 0.81 & 6.39 & 0.09 &  9\% \\
Disk & $-$1.99 & 0.14 & 13.35 & 1.45 & 7.18 & 0.23 & 57\% \\
Spiral A & $-$1.95 & 0.10 & 18.43 & 0.51 & 6.42 & 0.07 & 10\% \\
Spiral B & $-$2.01 & 0.10 & 18.78 & 0.63 & 6.48 & 0.08 & 11\% \\
\midrule
NGC1566 & $-$2.19 & 0.12 & 22.76 & 0.98 & 7.68 & 0.03 & 7.55  \\
\midrule
Central & $-$2.51 & 0.17 & 22.09 & 2.12 & 5.85 & 0.23 &  2\% \\
Disk & $-$1.78 & 0.11 & 15.31 & 1.42 & 7.48 & 0.22 & 84\% \\
Spiral A$_{in}$ & $-$1.62 & 0.11 & 26.11 & 0.98 & 6.10 & 0.08 &  4\% \\
Spiral B$_{in}$ & $-$1.47 & 0.10 & 25.49 & 0.82 & 6.00 & 0.07 &  3\% \\
Spiral A$_{out}$ & $-$1.78 & 0.11 & 20.74 & 1.06 & 6.08 & 0.12 &  3\% \\
Spiral B$_{out}$ & $-$2.17 & 0.13 & 18.76 & 1.22 & 6.19 & 0.16 &  4\% \\
\midrule
NGC3627 & $-$2.86 & 0.06 & 24.62 & 0.67 & 7.53 & 0.03 & 7.07  \\
\midrule
Central & $-$4.11 & 0.24 & 26.14 & 2.66 & 5.39 & 0.24 &  2\% \\
Bar & $-$2.12 & 0.12 & 36.47 & 1.11 & 5.15 & 0.06 &  1\% \\
Disk & $-$1.93 & 0.10 & 22.74 & 0.57 & 6.55 & 0.06 & 30\% \\
Spiral A & $-$1.85 & 0.11 & 19.50 & 1.45 & 6.60 & 0.18 & 34\% \\
Spiral B & $-$1.73 & 0.10 & 22.30 & 1.14 & 6.58 & 0.12 & 32\% \\
\bottomrule
\end{tabularx}
\end{table}


\section{Conclusions}
\label{sec:conclusion}

For the three late-type spiral galaxies under investigation---NGC~1097, NGC~1566, and~NGC~3627---we have obtained images from DustPedia~\cite{dustpedia} in seven infrared (IR) bands (3.6, 8, 24, 70, 100, 160, and~250~$\upmu$m). We performed photometric decomposition using two types of models: the S{\'e}rsic profile, as~previously applied in~\cite{Mosenkov2019}, and~a full multicomponent model. The~latter includes the explicit modeling of spiral arms, utilizing our previously developed method introduced in~\cite{Chugunov2024}.  

Consistent with our earlier studies of M\,51~\cite{Marchuk_M51} and HST/JWST data~\cite{Chugunov2025}, we confirm that the model that we developed can be effectively applied to analyze images across a broad range of wavelengths. Our findings include the following:  

\begin{enumerate}[label=(\roman*),leftmargin=7.5mm,labelsep=0.5mm,topsep=3pt]
\item The parameters of the spiral arms in the IR bands are similar to those obtained earlier for M\,51 and other galaxies. Specifically, the~spiral contribution to total luminosity ranges from 20\% to 60\%, increasing in the FIR bands. The~width of the arms is minimal at 24--70~$\upmu$m.  

\item Regarding dust distribution, we find that, in all cases, the~emission profile is well described by an exponential disk, provided that non-axisymmetric components such as spiral arms and bars are properly accounted for. This effect is most evident in NGC~3627, where the atypical dust distribution profile (S{\'e}rsic index \(n \sim 0\)) results from a bump induced by the influence of spiral arms. For~NGC~1097, the~large S{\'e}rsic index (\(n \gg 1\)) is attributed to the presence of a bright AGN and a nuclear ring. Although~these three cases do not fully resolve the question of the central dust depletion raised in~\cite{Mosenkov2019}, they illustrate how careful modeling in late-type galaxies can address this~issue. 

\item By utilizing bands associated with old stars (which trace the gravitational potential) and recent star formation, and~considering the full light distribution along the spiral arms, we tested the predictions of the stationary density wave theory. The~measured pitch angles are approximately 20$^{\circ}$, and~the angles in different wavelengths do not reveal expected inequalities, and thus do not support the presence of a stationary density wave. Furthermore, the~angular offsets between individual tracers also fail to align with this theory. Given the presence of companion galaxies in all cases, it is {more} likely that the spiral arms have a dynamic or tidal origin rather than being driven by a stationary density~wave.  

\item We performed SED fitting in the wavelength range of 3.6~$\upmu$m to 250~$\upmu$m for individual galaxy components, including spiral arms, which has not been performed in this manner before. Our results indicate that the average PAH fraction (\(q_{\mathrm{PAH}}\)) is nearly identical in the disk and spiral arms. Additionally, spiral arms contain a significant {(10\%-60\%)} fraction of the total dust in galaxies, with~an estimated dust temperature (\(T_{\mathrm{dust}}\)) of approximately 18--20~K, indicating that the dust is predominantly cold. Furthermore, we observe evidence of a dust temperature gradient in NGC~1566, where the separate modeling of the inner and outer spiral arms revealed a systematic variation in \(T_{\mathrm{dust}}\).  

\end{enumerate}

These results underscore the importance of incorporating spiral arms into photometric decomposition and the critical role that they play in addressing key questions about galaxy evolution. This pilot study serves as a foundation for future research and will be expanded to a larger sample of galaxies as additional observations become~available.




\vspace{6pt} 





\authorcontributions{Conceptualization, A.A.M. and I.V.C.; methodology, A.A.M., F.G., V.S.K., and I.V.C.; validation, A.A.M., I.V.C., F.G., and V.S.K.; investigation, A.A.M., V.S.K., and P.V.S.; data curation, A.A.M. and P.V.S.; writing---original draft preparation, A.A.M.; writing---review and editing, A.A.M., A.V.M., I.V.C., G.A.G., V.B.I., S.S.S., A.A.S., and D.M.P.; visualization, A.A.M. and I.V.C. All authors have read and agreed to the published version of the manuscript.}

\funding{We 
acknowledge financial support from the Russian Science Foundation, grant no.~20-72-10052.}

\dataavailability{The data underlying this article will be shared upon reasonable request to the corresponding author.}

\acknowledgments{DustPedia is a collaborative focused research project supported by the European Union under the Seventh Framework Programme (2007-2013) call (proposal no. 606847). The~participating institutions are Cardiff University, UK; National Observatory of Athens, Greece; Ghent University, Belgium; Université Paris Sud, France; National Institute for Astrophysics, Italy and CEA, France.}

\conflictsofinterest{The authors declare no conflicts of interest.} 


\newpage
\abbreviations{Abbreviations}{
The following abbreviations are used in this manuscript:\\

\noindent 
\begin{tabular}{@{}m{1.2cm}<{\raggedright}l}
SED & Spectral energy distribution\\
BIC & Bayesian information criterion\\
CR & Corotation radius\\
PSF & Point spread function\\
FWHM & Full width at half maximum\\
SPIRE & Spectral and Photometric Imaging REceiver\\
PACS & Photodetector Array Camera and Spectrometer\\
IRAC & InfraRed Array Camera\\
NIR & Near-infrared\\
MIR & Mid-infrared\\
FIR & Far-infrared
\end{tabular}
}

\appendixtitles{yes} 
\appendixstart
\appendix
\section[\appendixname~\thesection. Models for All Bands]{\large Models for All Bands}
\label{sec:appendix}

\begin{figure}[H]
\includegraphics[width=12.8cm]{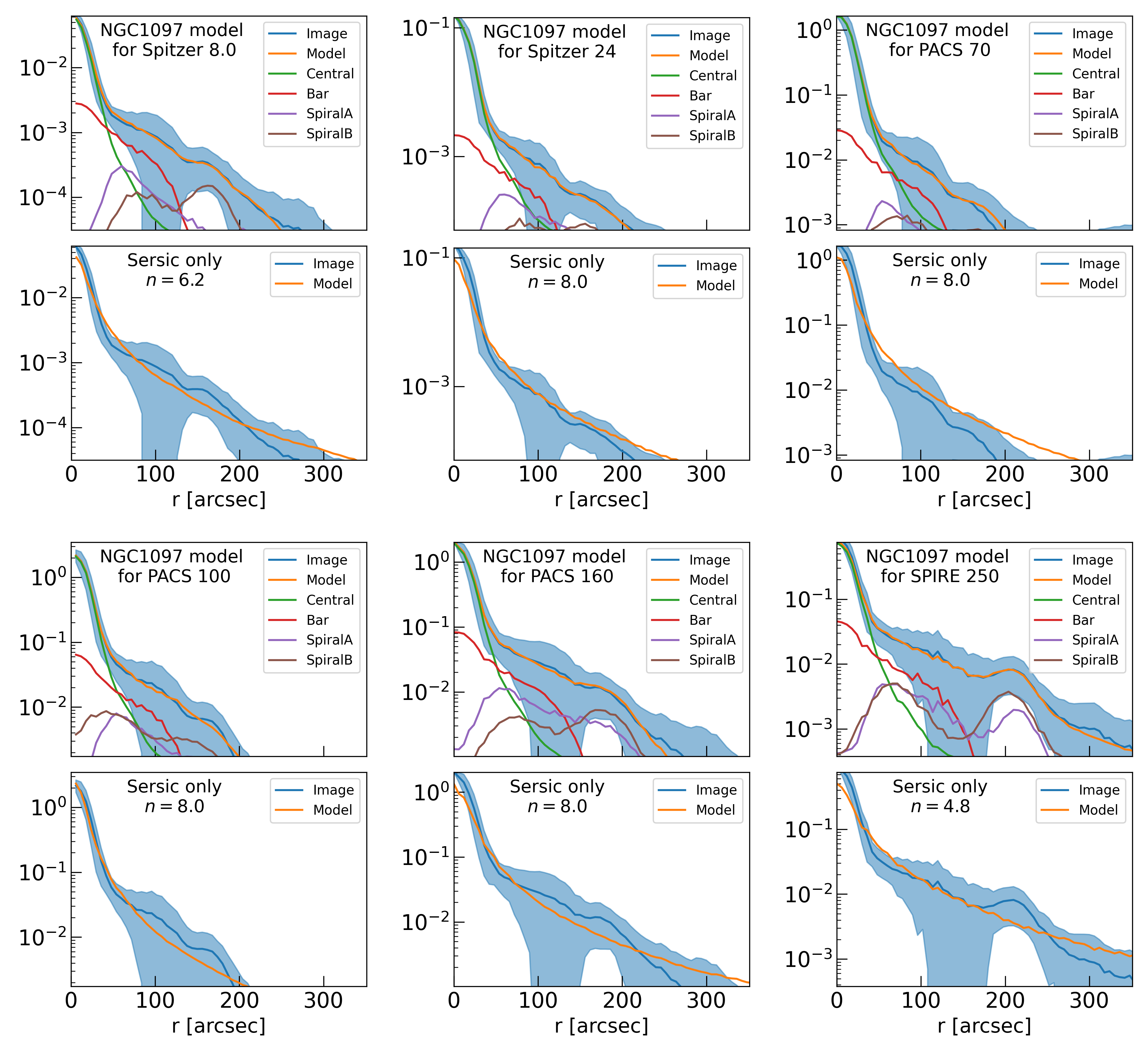}
\caption{Azimuthally averaged profiles for the NGC~1097 model and data in all bands except 3.6~$\upmu$m, which is presented in Figure~\ref{fig:2dcolumn}. In~each pair of images, the~full model is shown above the S{\'e}rsic model. The~vertical axis represents flux in Jy. The~dispersion of the observational data is indicated by the blue-filled area, with~the blue line representing the median. Each model component is shown as a line with its corresponding color. The~disk components are omitted for NGC~1097 to highlight their minor contribution to the model (see text for details). \label{fig:1097appendix}}
\end{figure}
\begin{figure}[H]
\includegraphics[width=12.8cm]{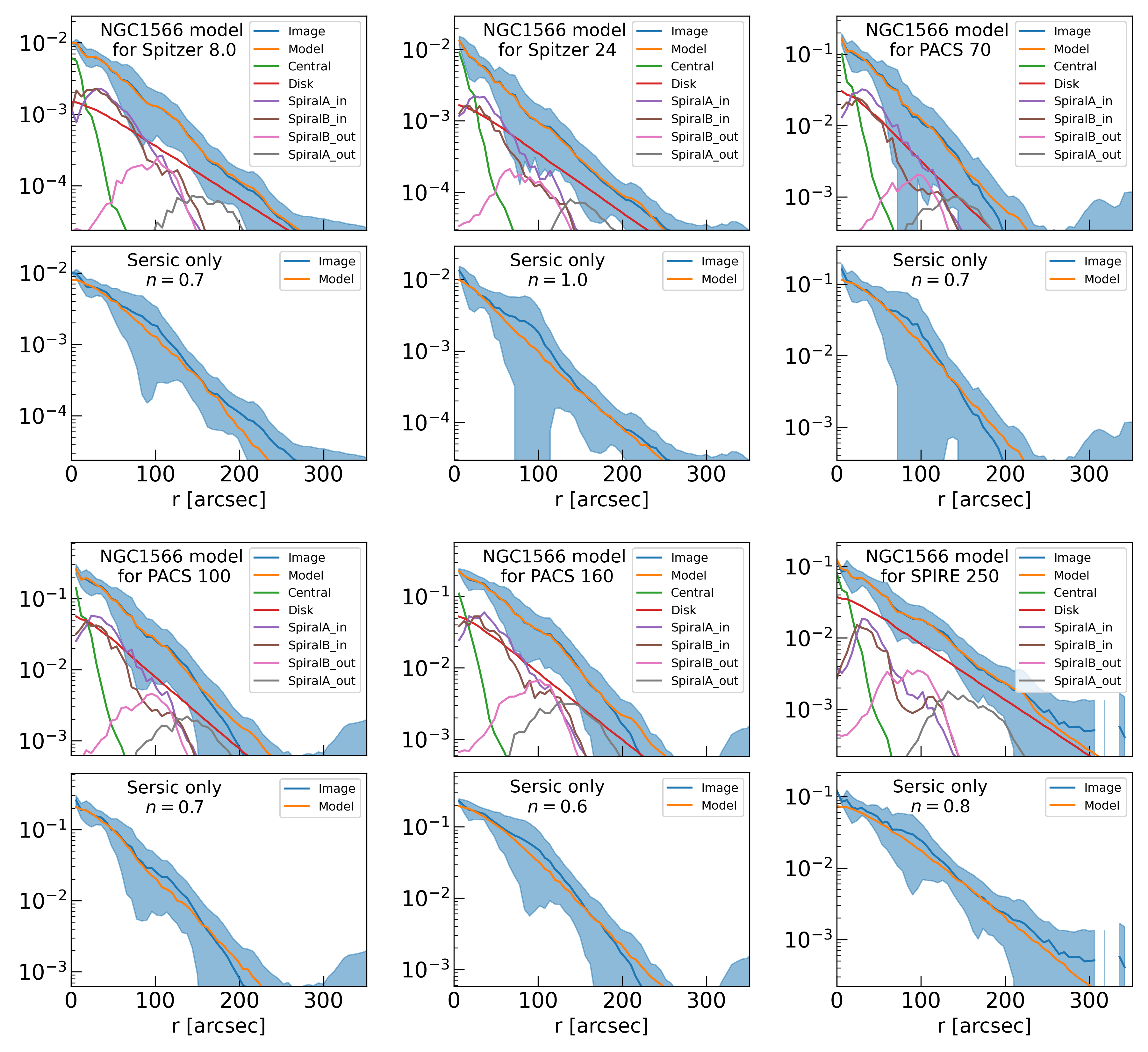}
\caption{Same as Figure~\ref{fig:1097appendix}, but~for NGC~1566.\label{fig:1566appendix}}
\end{figure}
\begin{figure}[H]
\includegraphics[width=12.8cm]{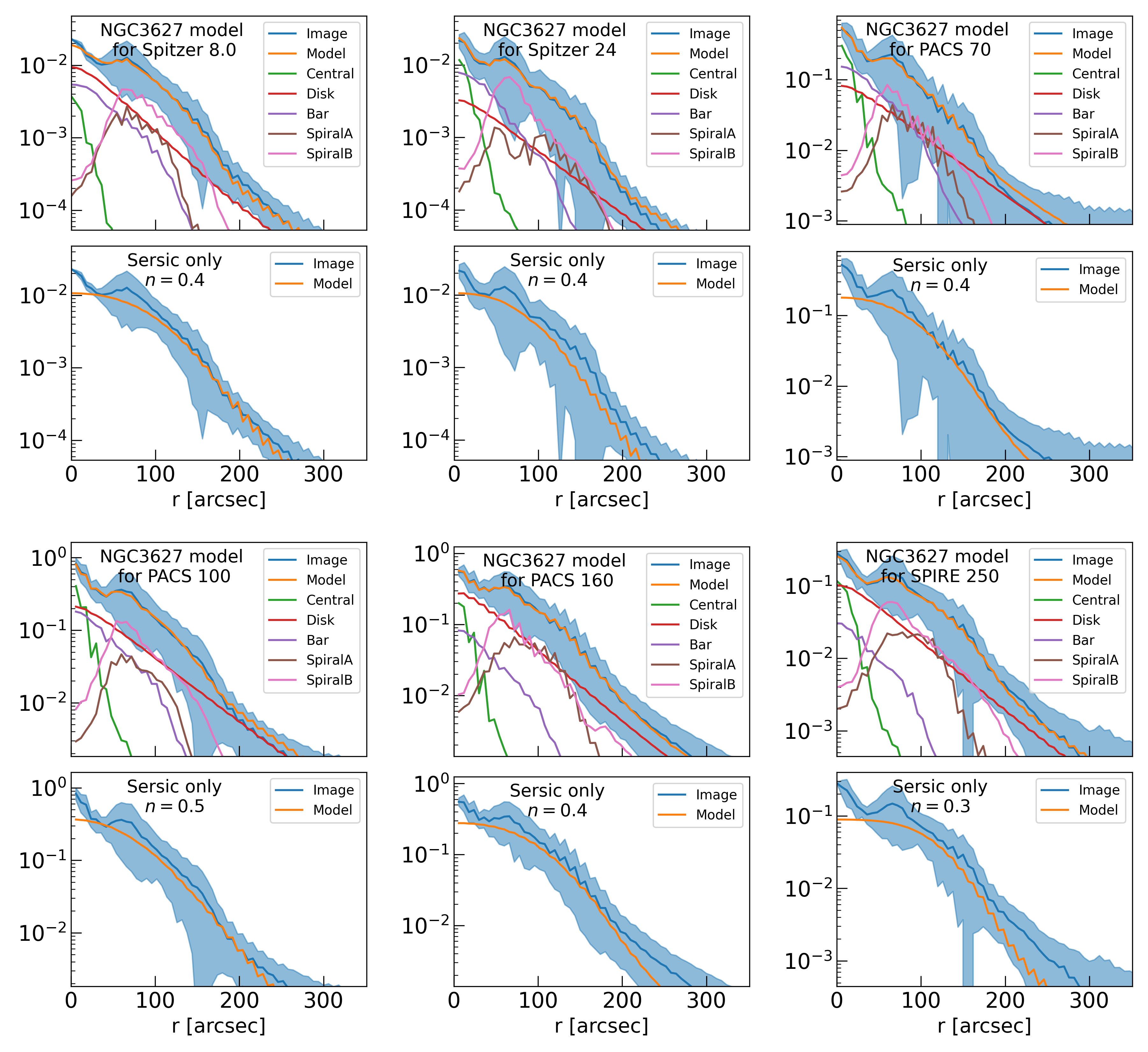}
\caption{Same as Figure~\ref{fig:1097appendix}, but~for NGC~3627. \label{fig:3627appendix}}
\end{figure}

This appendix presents data and models for six bands in the $8\upmu$m--$250\upmu$m range, shown in Figures~\ref{fig:1097appendix}--\ref{fig:3627appendix}. Models for the Spitzer 3.6~$\upmu$m images are displayed in \mbox{Figure~\ref{fig:2dcolumn}}. To~optimize space and provide an alternative perspective, we present azimuthally averaged~profiles.

\unskip   
\unskip   






\section[\appendixname~\thesection. HerBIE SED Fitting Example]{\large HerBIE SED Fitting Example}
\label{sec:sedappendix}
\vspace{-10pt}
\begin{figure}[H]
\includegraphics[width=12 cm]{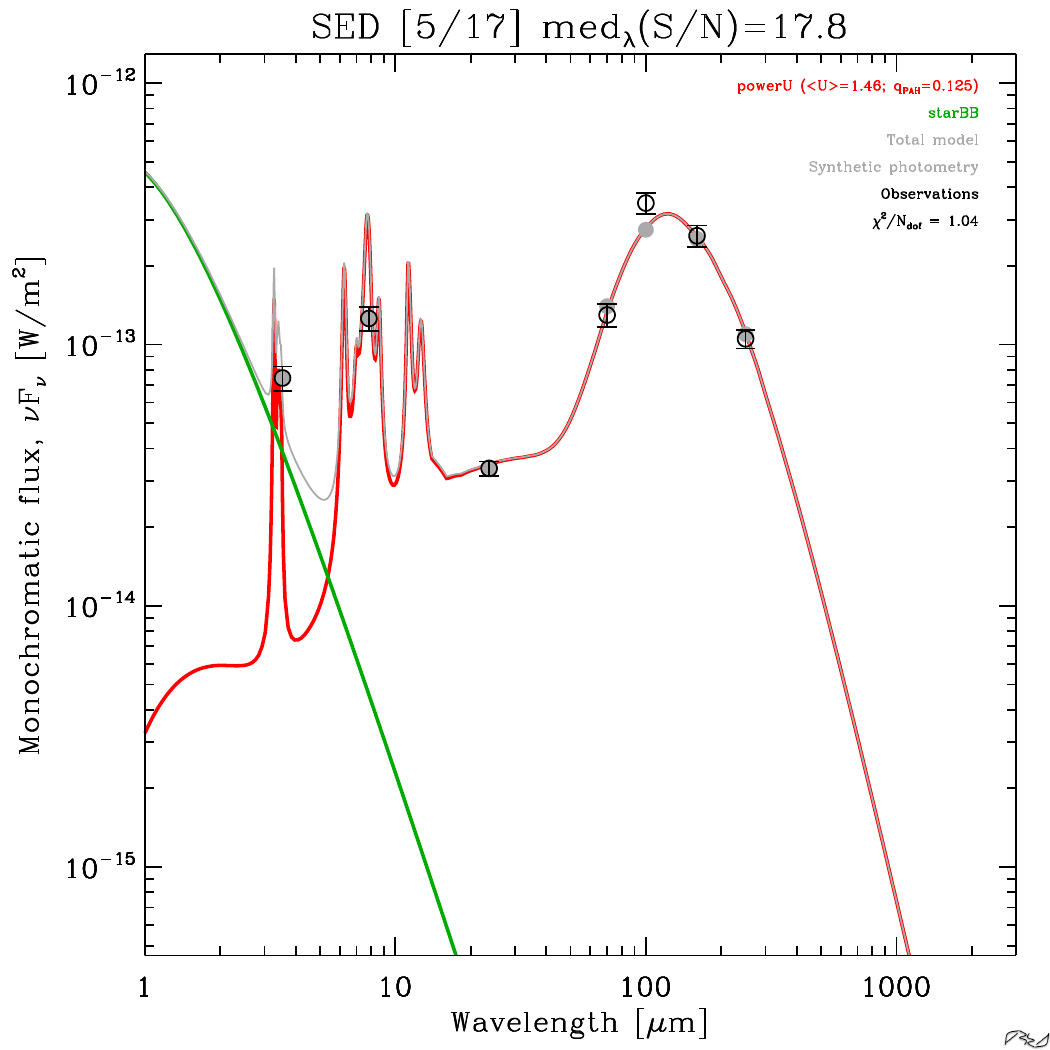}
\caption{Example 
of SED fitting using HerBIE for Spiral~B in NGC~1097. \label{fig:1566appendixsed}}
\end{figure}   

\begin{adjustwidth}{-\extralength}{0cm}
\printendnotes[custom] 

\reftitle{References}

\PublishersNote{}
\end{adjustwidth}
\end{document}